\documentclass[nofootinbib,aps,twocolumn,groupedaddress]{revtex4-1}

\usepackage{lipsum}
\usepackage{graphicx} 
\usepackage{amsfonts}
\usepackage{amsmath}
\usepackage{braket}
\usepackage{gensymb}
\usepackage{hyperref}
\usepackage{etoolbox}

\newcommand{\angstrom}{\textup{\AA}}

\makeatletter
\renewcommand\subsubsection{%
  \@startsection
    {subsubsection}%
    {3}%
    {\z@}%
    {.8cm \@plus1ex \@minus .2ex}%
    {.5cm}%
    {\normalfont\small\itshape\bfseries\centering}%
}%
\makeatother

\begin{document}


\title{$^{31}$P NMR study of discrete time-crystalline signatures in an ordered crystal of ammonium dihydrogen phosphate}


\author{Jared Rovny}
\author{Robert L. Blum}
\author{Sean E. Barrett} 
\email[]{sean.barrett@yale.edu}
\homepage[]{http://opnmr.physics.yale.edu/}
\affiliation{Department of Physics, Yale University, New Haven, Connecticut 06511, USA}


\date{\today}

\begin{abstract}
  The rich dynamics and phase structure of driven systems include the recently described phenomenon of the ``discrete time crystal'' (DTC), a robust phase which spontaneously breaks the discrete time translation symmetry of its driving Hamiltonian. Experiments in trapped ions and diamond nitrogen vacancy centers have recently shown evidence for this DTC order. Here we show nuclear magnetic resonance (NMR) data of DTC behavior in a third, strikingly different system: a highly ordered spatial crystal in three dimensions. We devise a DTC echo experiment to probe the coherence of the driven system. We examine potential decay mechanisms for the DTC oscillations, and demonstrate the important effect of the internal Hamiltonian during nonzero duration pulses.
\end{abstract}


\maketitle



\section{INTRODUCTION}  

In 2012, Wilczek proposed the existence of a system which spontaneously breaks time translational symmetry, dubbed a ``time crystal'' by analogy with a regular crystal, whose structure spontaneously breaks translational symmetry in space \cite{Wilczek2012}. While subsequent no-go theorems excluded the possibility of finding equilibrium states with this property \cite{Watanabe2015}, driven systems remained viable candidates. Multiple theoretical studies showed that driven systems could exhibit a rich phase structure, including a discrete time-crystalline (DTC) phase (also known as a Floquet time crystal, or a $\pi$-spin glass) \cite{Sacha2015, Sacha2018, Moessner2017}. For these driven systems, the time translation symmetry is discretized to the period of the drive, and the discrete time translation symmetry is broken by a state with oscillations at integer multiples of the drive period \cite{Khemani2016, vonKeyserlingk2016, Else2016}. However, this might be difficult to observe experimentally, since driven systems tend to thermalize as they absorb energy from the drive, which could prevent the experimental observation of DTC signatures \cite{DAlessio2014, Lazarides2014, Lazarides2014a, Lazarides2015, Lazarides2017}. To avoid this fate, many DTC models worked in a regime that favored many-body localization (MBL) \cite{Khemani2016, vonKeyserlingk2016, vonKeyserlingk2016a, vonKeyserlingk2016b, Moessner2017}; other models predicted that the DTC could be observed without MBL, in a prethermal regime \cite{Abanin2015, Else2016, Else2017, Machado2017, Kuwahara2016, Abanin2017, Abanin2017a}.
 
After Yao \textit{et al}. \cite{Yao2017} proposed experimental realizations, evidence for DTC order was obtained in two very nice experiments: one using trapped ions \cite{Zhang2017} and the other using diamond nitrogen vacancy (NV) centers \cite{Choi2017}. The experiment in trapped ions was closer to original theoretical models for DTC order, and included elements more conducive to MBL, such as a one-dimensional spin chain with $\sim$10 spins, spin-spin interactions that fell off as $\sim$$r^{-1.51}$, and high-variance on-site disorder \cite{Zhang2017}. The experiment in diamond NV centers \cite{Choi2017} was strikingly different from theoretical models, especially in that it used a three-dimensional system of spins at random locations, with spin-spin interactions that fell off as $\sim$$r^{-3}$; these characteristics are expected to preclude MBL \cite{Oganesyan2007, Huse2014, Nandkishore2015, Ho2017}. While disorder did exist in the system of diamond NV centers, followup studies have proposed alternatives to MBL as mechanisms for the observed signatures of DTC order \cite{Ho2017, Huang2017, Russomanno2017}.

 In this paper, we report the observation \cite{Rovny2018} of signatures of a DTC in an ordered spatial crystal even further from ideal MBL conditions than all prior DTC experiments. We also study the lifetime of the DTC oscillations, demonstrating that a significant part of the observed decay envelope is due to coherent evolution. Finally, we describe the way in which the lifetime of the observed DTC oscillations strongly depends on the action of the internal Hamiltonian during an applied pulse; we demonstrate control of this decay mechanism, which may be important for experiments which strive to observe the intrinsic lifetime of the DTC.

 \section{PHYSICAL CHARACTERISTICS OF THE SYSTEM}

\begin{figure}
  \includegraphics{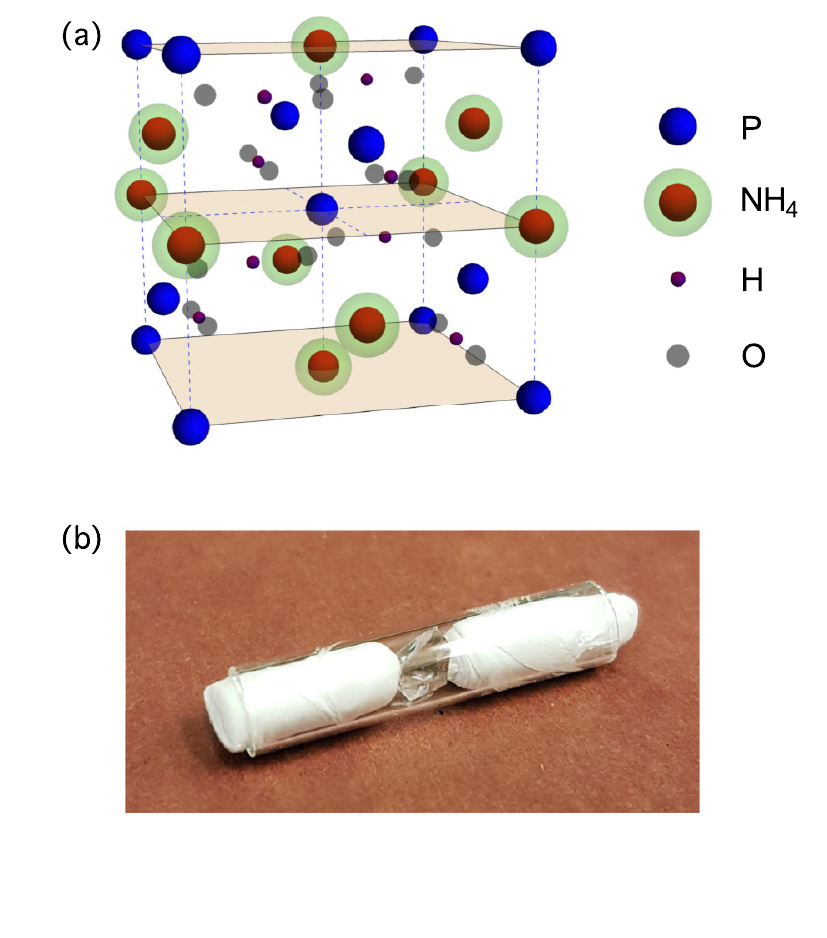}%
\caption{\label{ADPStructure}(a) Atoms in the unit cell of ammonium dihydrogen phosphate (ADP), which has chemical formula NH$_4$H$_2$PO$_4$. ADP is an ionic, tetragonal crystal with space group $I\overline{4}2d$. At room temperature, the NH$_4$ groups experience rapid in-place rotation, such that the time-averaged location of the four $^1$H is at the nitrogen site. The $^1$H in NH$_4$ are shown in a distributed manner to reflect this. We also place the remaining so-called ``acid'' protons ($^1$H) in their time-averaged positions, between the lattice sites of the nearest oxygens \cite{West1930, Khan1973}. (b) The ADP crystal sample studied here, shown in a 5-mm-diameter NMR sample tube, held in place by rolled teflon tape (white).}
\end{figure}

 In this section, we discuss the methods used in characterizing the system of $^{31}$P nuclear spins in ammonium dihydrogen phosphate (ADP), and present the key features of the system and its internal spin Hamiltonian. We begin with an overview of nuclear magnetic resonance (NMR) methods and useful terminology, then discuss the application of these methods to the particular system of $^{31}$P nuclear spins in an ADP crystal (Fig. \ref{ADPStructure}).

\subsection{NMR overview}

Our experiments are carried out at room temperature in the presence of a strong ($H_0~=~4$\,T) external static magnetic field. Thus, we can use the strong-field, high temperature approximation to write the equilibrium density matrix for the nuclear spins; to calculate the detected signal, it is sufficient to start with the ``reduced'' density matrix $\rho^\text{lab}_0=I_{z'_T}$, where we have taken $H_0$ to be in the $z'$ direction \cite{Slichter1996, Mehring1983}. These nuclear spins precess around the strong external field at their Larmor frequencies $\omega_0=\gamma H_0$, where $\gamma$ is the gyromagnetic ratio for the spin species; in NMR, the observable signal is the voltage induced in a detection coil by the time-varying flux arising from the precessing nuclear spin magnetization, $\braket{M_{y'_T}(t)}=\gamma\hbar \braket{I_{y'_T}(t)}=\gamma\hbar\text{Tr}[I_{y'_T}\rho^\text{lab} (t)]$, where $y'$ is the axis of the coil. The time evolution operator $\mathcal{U}^\text{lab}(t;0)$, which determines $\rho^\text{lab}(t)=\mathcal{U}^\text{lab}(t;0)\rho_0^\text{lab}\mathcal{U}^\text{lab}(t;0)^{-1}$, is itself determined by the relevant Hamiltonian \cite{Slichter1996}, which can in general be time-dependent.

In the laboratory frame (in the absence of applied pulses), the spin Hamiltonian is $\mathcal{H}^{\text{lab}}=\mathcal{H}_0 + \mathcal{H}^{\text{lab}}_{\text{int}}$, where the scale of the term due to the static external field, $\mathcal{H}_0=- \hbar \omega_0 I_{z'_T}$, is 4 to 5 orders of magnitude larger than the scale of any terms in the internal spin Hamiltonian $\mathcal{H}^{\text{lab}}_{\text{int}}$. Thus, we may write the secular internal Hamiltonian $\mathcal{H}_{\text{int}}$ in the frame that is rotating about $z'$ at the Larmor frequency $\omega_0$, ignoring terms which are nonsecular in the rotating frame (to a very good approximation). The rotating frame axes are $(x,y,z)$, where $z\parallel z'$, so $\rho_0 = \rho^\text{lab}_0$.

To manipulate the nuclear spins, we apply strong radiofrequency (rf) pulses at the Larmor frequency of the particular spin species to be manipulated (see Table \ref{SpinsTable}). For the duration of an applied pulse, the rotating frame Hamiltonian becomes $\mathcal{H_P}=\mathcal{H}_{\text{int}}+\mathcal{H}_{\text{rf}}$, with the added external term $\mathcal{H}_{\text{rf}}= -\hbar \omega_1 I_{\phi_T}$ for a pulse of strength $\omega_1$ and phase $\phi$. To calibrate $\omega_1$ for a given pulse power, we use a nutation experiment \cite{Slichter1996}. The pulses are applied for duration $t_p$, such that for e.g. a pulse of angle $\pi$, we have $\omega_1 t_p = \pi$. Because $\omega_1$ is typically large for applied pulses (for instance, $\omega_1 / 2\pi \approx 68$\,kHz in our experiment), $\mathcal{H}_{\text{int}}$ is usually ignored for the duration of the pulse (the delta-function pulse approximation) \cite{Slichter1996, Abragam1961, Mehring1983, Ernst1987, Freeman1997} --- we will revisit this approximation below. In this paper, we will use the symbol $\phi_{\theta}$ to represent a pulse of angle $\theta$ applied at phase $\phi$, emphasizing the phase of the pulse. 

The basic NMR experiment measures a free-induction decay (FID) by applying a $\theta = \pi/2$ pulse to spins starting with an equilibrium $z$-magnetization, to produce measurable magnetization along $\hat{y}$: $\{X_{\pi/2} - \text{FID}\}$, where FID represents the acquisition of the signal as a function of time after the first pulse \cite{Slichter1996}. A Fourier transform (FT) of the resulting time data $\braket{I_{y_T}(t)}$ yields a line shape for the observed spins, which reflects the action of the full $\mathcal{H}_{\text{int}}$. To remove the effect of Zeeman terms in the internal Hamiltonian, a Hahn echo sequence may instead be used, which includes a $\pi$ pulse between the preparation pulse and the final readout to ``refocus'' the Zeeman-dephased signal into an echo: $\{X_{\pi/2} - \tau - Y_{\pi} - \tau - \text{Echo}\}$ \cite{Hahn1950}. The echo amplitude measured as a function of $\tau$ can be used to create a ``pseudo-FID''; the corresponding spectrum reflects the unrefocused parts of $\mathcal{H}_\text{int}$. Each of these pulse sequences will be used in characterizing the system below.

Aside from these, we will use two further techniques common in NMR: cross polarization and spin decoupling. The first, cross polarization (CP), takes advantage of the higher polarization that exists in equilibrium spin ensembles with higher gyromagnetic ratios $\gamma_S$, using it as a source to augment the lower polarization of the measured, target spins (with $\gamma_I$). To accomplish this, rf fields $H^S_1,H^I_1$ are applied at the Larmor frequencies of the two spins to be cross polarized, such that the effective Zeeman energy levels are equalized in the tilted, doubly rotating frame (the ``Hartmann-Hahn'' matching condition \cite{Hartmann1962}): $\gamma_SH^S_1=\gamma_IH^I_1$. While this can be used to boost the polarization of the initial reduced density matrix up to $\rho_0'=(\gamma_S/\gamma_I)\rho_0$, an even more important benefit is that CP experiments on the target spins $I$ may be repeated on the much faster timescale of the source spins $S$ (for repolarization times $T_1^S \ll T_1^I$) \cite{Pines1972, Pines1973}. The second technique, spin decoupling, allows us to selectively remove the dipolar coupling between two spin species, by applying strong continuous-wave (cw) rf irradiation at the Larmor frequency of one of the spins \cite{Slichter1996}. The details of these techniques in our system will be discussed further below.

\subsection{NMR of $^{31}$P in ADP}

We study the ionic crystal ammonium dihydrogen phosphate [ADP, also called monoammonium phosphate (MAP)], with chemical formula NH$_4$H$_2$PO$_4$. We grew an ADP crystal by slow evaporation from aqueous solution [Fig. \ref{ADPStructure}(b)].  Simulations of the NMR spectra (discussed below) are consistent with our sample being a single crystal of a known orientation. This sample was being used as a test bed for controlling the $^{31}$P-$^1$H spin Hamiltonian in other materials; however, since both a sample and a double-resonance NMR system were available, we decided to try the DTC pulse sequence on ADP.

\subsubsection{$^{31}$P spin Hamiltonian in ADP}

ADP contains the nuclear spins summarized in Table \ref{SpinsTable}, but our analysis assumes that only $^{31}$P, $^1$H, and $^{14}$N are present (each at 100\% natural abundance). In our NMR experiments, we will detect the signal from the $^{31}$P spins. The Zeeman interaction of the $^{31}$P spins with the applied magnetic field $H_0=4$\,T dominates the spin Hamiltonian in the laboratory frame. Jumping to the frame rotating at the Larmor frequency of the $^{31}$P nuclei $(\omega_0 = \gamma_\text{P} H_0=2\pi\times 68.940\text{\,MHz})$, the secular terms in the internal spin Hamiltonian $\mathcal{H}_{\text{int}}$ for $^{31}$P include Zeeman interactions $\mathcal{H}_\text{Z}$, dipolar couplings among the same spin species (homonuclear, $\mathcal{H}_{zz}^\text{P,P}$), and dipolar couplings between different spin species (heteronuclear, $\mathcal{H}_{zz}^\text{P,H}$, $\mathcal{H}_{zz}^\text{P,N}$):

\begin{align}\label{Hamiltonian}
  \mathcal{H}_{\text{int}} &= \mathcal{H}_\text{Z}^\text{P} + \mathcal{H}_{zz}^\text{P,P} + \mathcal{H}_{zz}^\text{P,H} + \mathcal{H}_{zz}^\text{P,N} \nonumber \\
  &= \sum_{i}\Omega_i I_{z_i}
  +  \sum_{i,j>i}B^\text{P}_{ij} ( 3I_{z_i}I_{z_j} - \vec{I_i}\cdot\vec{I_j} ) \nonumber \\
  & + \sum_{i,j}B^\text{H}_{ij} (2I_{z_i}S_{z_j})
  + \sum_{i,j}B^\text{N}_{ij} (2I_{z_i}R_{z_j}).
\end{align}

Here, the coupling constants $B^\text{P}_{ij},B^\text{H}_{ij},$ and $B^\text{N}_{ij}$ are defined for the coupling of $^{31}$P to $^{31}$P, $^1$H, and $^{14}$N respectively. The coupling constant between a $^{31}$P spin $i$ and a spin $j$ (of spin species $\alpha=\{^{31}$P$, ^1$H$,^{14}$N$\}$) is

\begin{equation}\label{Bij}
   B^\alpha_{ij} =\frac{\mu_0}{4\pi}\frac{\gamma_\text{P}\gamma_\alpha \hbar^2}{|\vec{r}_{ij}|^3}\frac{1-3 \textrm{cos}^2(\theta_{ij})}{2}\,\text{,}
\end{equation}

where $\theta_{ij}$ is the angle between the internuclear vector $\vec{r}_{ij}$ and the $z$-axis (defined by the static external field), $\mu_0$ is the vacuum permeability, and $\gamma_\text{P}$ and $\gamma_\alpha$ are the nuclear gyromagnetic ratios for $^{31}$P and $\alpha$. $\{I_{\phi}, S_{\phi}, R_{\phi}\}$ are the spin operators for $\{^{31}$P$, ^1$H$,^{14}$N$\}$, with $\phi=x,y,z$ \cite{Slichter1996, Levitt2008}. Because the $^{31}$P sites in a single crystal are magnetically equivalent \cite{Eichele1994}, any variations in the Zeeman interaction will be small and slowly varying across the sample, arising from the sample's magnetic susceptibility or variations in the static external field --- in this sample, variations in the Zeeman interaction are less than 1 ppm relative to the static field (see next section). For this reason, we may replace the Zeeman Hamiltonian term with $\sum_{i}\Omega_i I_{z_i} \rightarrow \Omega_T I_{z_T}$ for any cluster of spins small relative to the size of the sample, where $\Omega_T/2\pi$ may be up to a few hundred Hz at most, caused by a resonance offset as the strong external field drifts slowly over the course of days or weeks. Because this Zeeman term has negligible variations from one spin site to the next (unlike most prior DTC models), this Hamiltonian retains unsuppressed ``flip-flop'' terms $I_{x_i}I_{x_j}+I_{y_i}I_{y_j}=(I^+_{i}I^-_{j}+I^-_{i}I^+_{j})/2$ for the homonuclear dipolar coupling, as well as long-range Ising-type couplings to $^1$H and $^{14}$N. Another feature of our experiment is that the coupling to the $^1$H can be selectively ``turned off'' with high-power cw decoupling at the $^1$H Larmor frequency \cite{Mehring1983}, which we will refer to as ``$^1$H off.'' We will refer to experiments that do not use cw decoupling as ``$^1$H on.''

A further type of order arises from the symmetries of the ADP crystal itself (see Appendix \ref{CrystalSymmetry} for details): the particular symmetry of the $^{31}$P and $^{14}$N sublattices leaves the set of geometric factors $B^\text{P}_{ij}$ and $B^\text{N}_{ij}$ invariant from one $^{31}$P site to the next, for each $i$. The coupling constants $B^\text{H}_{ij}$ do not obey the same symmetry except for certain ``special'' crystal orientations relative to the external field; in general, there are two distinct sets of $B^\text{H}_{ij}$ for a given $i$, which become the same at the crystal orientation that is consistent with our measured NMR spectra (see below).

\begin{table}
  \caption{\label{SpinsTable} Spins present in ADP, with their Larmor frequencies $\omega_0/2\pi$ in the presence of a strong $H_0=4$\,T magnetic field. In our analysis, we ignore the presence of the rare $^2$H, $^{15}$N, and $^{17}$O nuclear spins.}
\begin{ruledtabular}
  \begin{tabular}{l c c r}
  Nuclide & Natural abundance & Spin & $\omega_0 / 2\pi$ at 4 T (MHz) \\
  \hline
  $^{1}$H & 99.98\% & 1/2 & 170.304  \\
  $^{31}$P & 100\% & 1/2 & 68.940  \\
 $^{14}$N & 99.64\% & 1 & 12.307  \\
  $^2$H & 0.02\% & 1 & 26.143  \\
  $^{15}$N & 0.37\% & 1/2 & -17.265  \\
  $^{17}$O & 0.04\% & 5/2 & -23.093 
  \end{tabular}
\end{ruledtabular}
\end{table}

\begin{figure}
\includegraphics[width=0.5\textwidth]{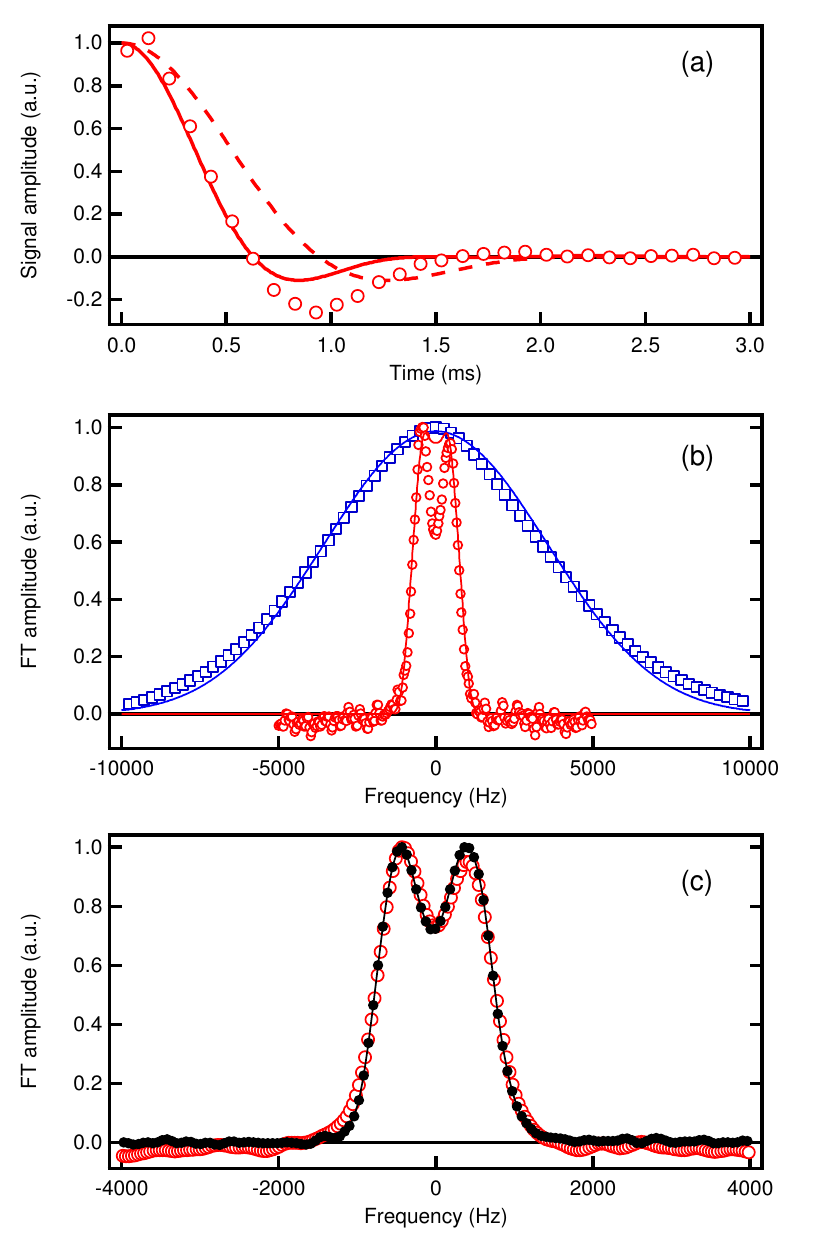}%
\caption{\label{LineShape}(a) Magnetization decay from a Hahn echo experiment with $^1$H off (circles), where each data point is acquired with a Hahn echo sequence for a different value of $\tau$. We compare this to the simulated decay from an Ising-type approximation, both before (dashed line) and after (solid line) scaling $B_{ij}$ by $3/2$ to approximate the actual dipolar Hamiltonian. (b) $^{31}$P spectra as acquired by an FID with $^1$H on (blue squares), and by a Hahn echo with $^1$H off [red circles, FT of Hahn echo data in (a)], with the results of a numerical model at a single crystal orientation (lines). (c) Comparison of the $^{31}$P spectrum from an FID (closed circles) to the line shape from an altered Hahn echo (open circles). The Hahn echo spectrum has been broadened using a Gaussian with FWHM 280 Hz, to account for the $^{31}$P-$^{14}$N coupling.}
\end{figure}

\subsubsection{Simulating the observed spectra}

To verify our understanding of the crystal structure and orientation, we compare simulations of the dipolar line shapes to data from Hahn echo and FID experiments. First, we measure the $^{31}$P-$^{31}$P dipolar line shape (from $\mathcal{H}_{zz}^\text{P,P}$) using a Hahn echo experiment with $^1$H off [Fig. \ref{LineShape}(a), open circles]: the decoupling removes the effect of $\mathcal{H}_{zz}^\text{P,H}$, and we expect the $\pi$ pulses of the Hahn echo to refocus (and thus remove the effects of) $\mathcal{H}_{zz}^\text{P,N}$ and $\mathcal{H}_Z^\text{P}$. To simulate this $^{31}$P-$^{31}$P line shape (see Appendix \ref{Numerics} for details), we start with the exactly-solvable Ising-type approximation for the dipolar coupling between ``unlike'' spins $I$ and $S$: $\mathcal{H}_{\text{Ising}}=\sum_{i,j}B_{ij}(2I_{z_i}S_{z_j})$, which produces the dashed line in Fig. \ref{LineShape}(a). The dashed line fails to describe the data [Fig. \ref{LineShape}(a), open circles], which conforms to our expectation that the $^{31}$P-$^{31}$P coupling is really between ``like'' spins.  Unfortunately, an exact treatment of the signal decay for ``like'' spins requires a full density matrix calculation; our dense lattice of $^{31}$P spins is hard to model accurately in the typical limit of $N<10$ spins \cite{Li2008}.  Instead, we try to approximate the ``like'' spin decay curve by a simple modification of the ``unlike'' spin curve. To approximate the actual $I_{z_i}I_{z_j}$ coefficient in the full dipolar coupling for ``like'' spins $\sum_{i,j>i} B_{ij}(3I_{z_i}I_{z_j}- \vec{I_{i}} \cdot \vec{I_{j}})$, we use the same analytic expression as in the ``unlike'' spin case, but with the $B_{ij}$ frequencies scaled up by $3/2$ \cite{Li2007,Li2008}. This produces the solid line in Fig. \ref{LineShape}(a), which lies very close to the Hahn echo data from our experiments. This Ising-type approximation produces a smaller oscillation in the time domain than the data exhibits [Fig. \ref{LineShape}(a)], creating a shallower dip at the center of the resulting spectrum than seen in the data [Fig. \ref{LineShape}(b), red line versus open circles]; similar results were seen in earlier uses of this approximation \cite{Li2008}.


Next, we study the full effect of $\mathcal{H}_\text{int}$ [Eq. (\ref{Hamiltonian})] by acquiring an FID with $^1$H on [spectrum in Fig. \ref{LineShape}(b), blue squares]. We simulate this spectrum by combining the separately calculated line shapes from $\mathcal{H}_{zz}^\text{P,P}$, $\mathcal{H}_{zz}^\text{P,H}$, and $\mathcal{H}_{zz}^\text{P,N}$. Each dipolar interaction is calculated using $\mathcal{H}_\text{Ising}$ (where the scaling by $3/2$ is only applied for the homonuclear $\mathcal{H}_{zz}^\text{P,P}$), and they are combined by multiplication in the time domain (see Appendix \ref{Numerics}). The final simulated spectrum is shown in Fig. \ref{LineShape}(b) (blue solid line), and is quite close to the measured spectrum.

Finally, we can study the Zeeman interaction, $\mathcal{H}_\text{Z}$, by comparing the spectra from both a Hahn echo and an FID with $^1$H off. We expect the difference between the these two spectra to arise only from $\mathcal{H}_\text{Z}$ and $\mathcal{H}_{zz}^\text{P,N}$, both of which are refocused in a Hahn echo, but not in an FID. In order to isolate the effect of $\mathcal{H}_\text{Z}$, we can ``put back'' the effect of $\mathcal{H}_{zz}^\text{P,N}$ into the Hahn echo spectrum using Gaussian line broadening, such that any remaining difference between the spectra is primarily attributable to $\mathcal{H}_\text{Z}$. We broaden the Hahn echo spectrum using a Gaussian with the same full width at half maximum (280 Hz) as the simulated spectrum for $\mathcal{H}_{zz}^\text{P,N}$. The resulting ``altered Hahn echo'' spectrum [Fig. \ref{LineShape}(c), red open circles] is very close to the FID spectrum [Fig. \ref{LineShape}(c), black closed circles], putting a small upper bound on the Zeeman spread in our system (estimated at 1 ppm relative to $H_0$).

 The orientation of the ADP crystal relative to the external field has a significant effect on the shape of the $^{31}$P spectrum \cite{Eichele1994}. We explore many possible crystal orientations in the simulations discussed in this section, where each simulation is calculated at a single crystal orientation. We parametrize the crystal orientation by the polar and azimuthal angles $(\theta_c, \phi_c)$ of the strong static field relative to the crystal axes $(a,b,c)$. We find the best agreement with the measured spectra [Figs. \ref{LineShape}(a) and \ref{LineShape}(b)] for orientation angles $(\theta_c,\phi_c) = (60^{\circ},0^{\circ})$ [we see identical results for $(\theta_c,\phi_c) = (n60^{\circ},m90^{\circ})$, with $n=\{1,2\}$, and $m=\{0,1,2,3\}$, because of the crystal symmetry]. Combined with the visual clarity of the crystal [Fig. \ref{ADPStructure}(b)], the agreement of the simulated spectra with experiment shown in Fig. \ref{LineShape} indicates that our sample is a single large crystal domain, and that we have a quantitative understanding of its spin Hamiltonian. From the numerics for the crystal orientation best matching the data, we can estimate the typical coupling strengths of the dipolar interactions as the root mean square (rms) angular frequencies $W^\text{P,H}$, $W^\text{P,P}$, and $W^\text{P,N}$ for the coupling of phosphorus to $^1$H, $^{31}$P, and $^{14}$N, respectively. We find $W^\text{P,H}/2\pi = 3500$\,Hz, $W^\text{P,P}/2\pi= 508$\,Hz, and $W^\text{P,N}/2\pi = 97$\,Hz, which added in quadrature give $W^\text{P,HPN}/2\pi= 3538$\,Hz (see Appendix \ref{Numerics}).

 \begin{figure}
\includegraphics{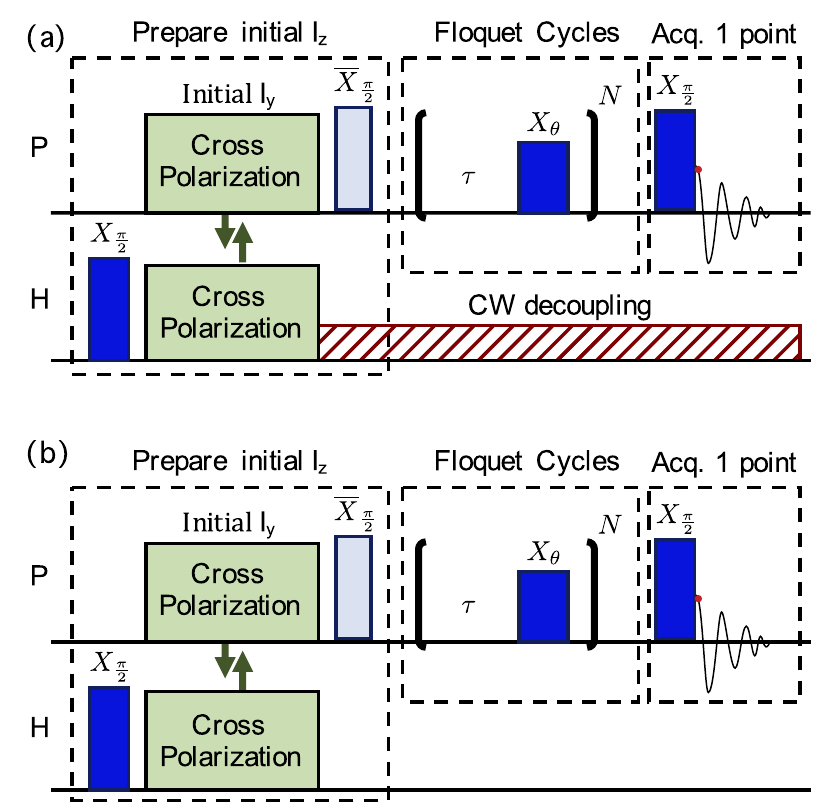}%
\caption{\label{DTCSequence}DTC pulse sequence. After $^1$H spins are rotated with an $X_{\pi/2}$ pulse (tall blue block), $^{31}$P magnetization is created along $\hat{y}$ via cross polarization with the $^{1}$H spins, and is then rotated into $\hat{z}$ with an $\overline{X}_{\pi/2}$ pulse (tall orange block) to prepare the initial state of the system. We then apply repeated Floquet cycles consisting of a delay $\tau$ followed by a pulse $X_\theta$ (wide blue block). After $N$ cycles, an $X_{\pi/2}$ pulse is applied, and the magnetization is immediately measured, producing a single data point in $S(t)$. We increase $N$ by 1 and repeat the sequence, following a 3-s recycle delay. This sequence is applied for $N=1,\! 2,\! \text{...},\! 128$. After cross polarization, continuous rf decoupling (red) can be applied ($^1$H off) to remove the effect of the $^1$H (a), or decoupling can be omitted ($^1$H on), allowing the $^1$H to act on the $^{31}$P spins (b).}
 \end{figure}

 \begin{figure*}
   \includegraphics{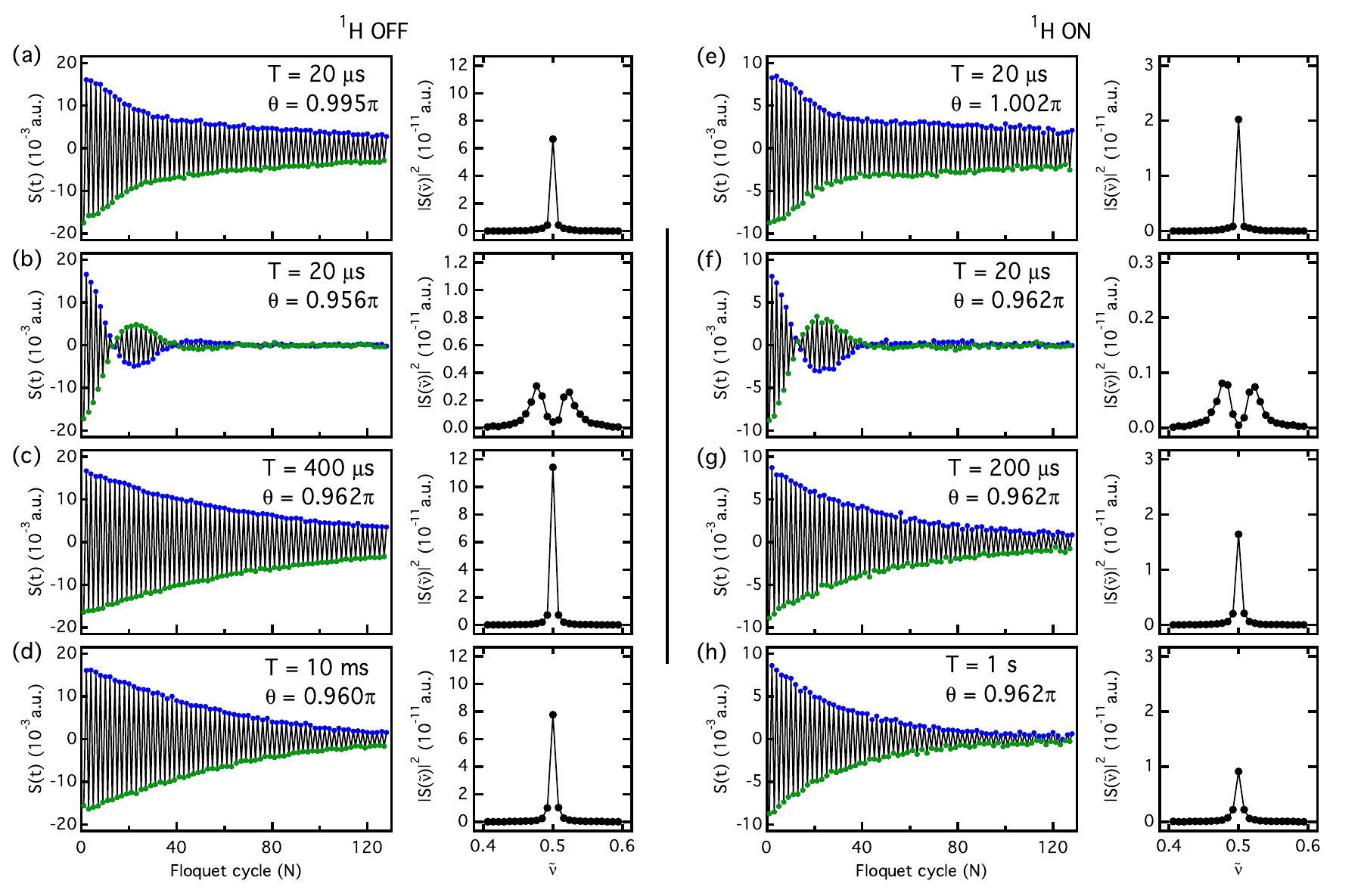}%
\caption{\label{CompareToLukin}(a) Applying repeated $\pi$ pulses with $^1$H decoupling at $\theta \approx \pi$ and small $\tau$, we see oscillations in the time-domain signal, corresponding to a single peak in the FT signal at $\tilde{\nu}=1/2$. Each data point is acquired in a separate experiment, using the DTC sequence for a given $N$. $N$-odd are in green (starting negative) and $N$-even are in blue (starting positive), with black lines between them to guide the eye. (b) Decreasing $\theta$, we observe beating in the time domain signal, corresponding to a splitting of the Fourier peak. (c) Near the same $\theta \approx 0.962 \pi$ but for increased $\tau$, the oscillations are restored, once again producing a single peak in the Fourier spectrum. (d) Significantly increasing $\tau$, we still see the same behavior.  (e)-(h) We observe qualitatively similar behavior in the absence of $^1$H decoupling. Note that for (h), $NT$ becomes comparable to the $^{31}$P lattice relaxation time, $T_1^\text{P}=103$\,s. In each case (a)-(h), $T=\tau + t_p$ with $t_p=7.5$\,$\mu$s. Data in (a)-(d) were acquired with 2$\times$ the number of scans as (e)-(h), doubling the maximum value of $S(t)$.}
\end{figure*}

\begin{figure}
\includegraphics{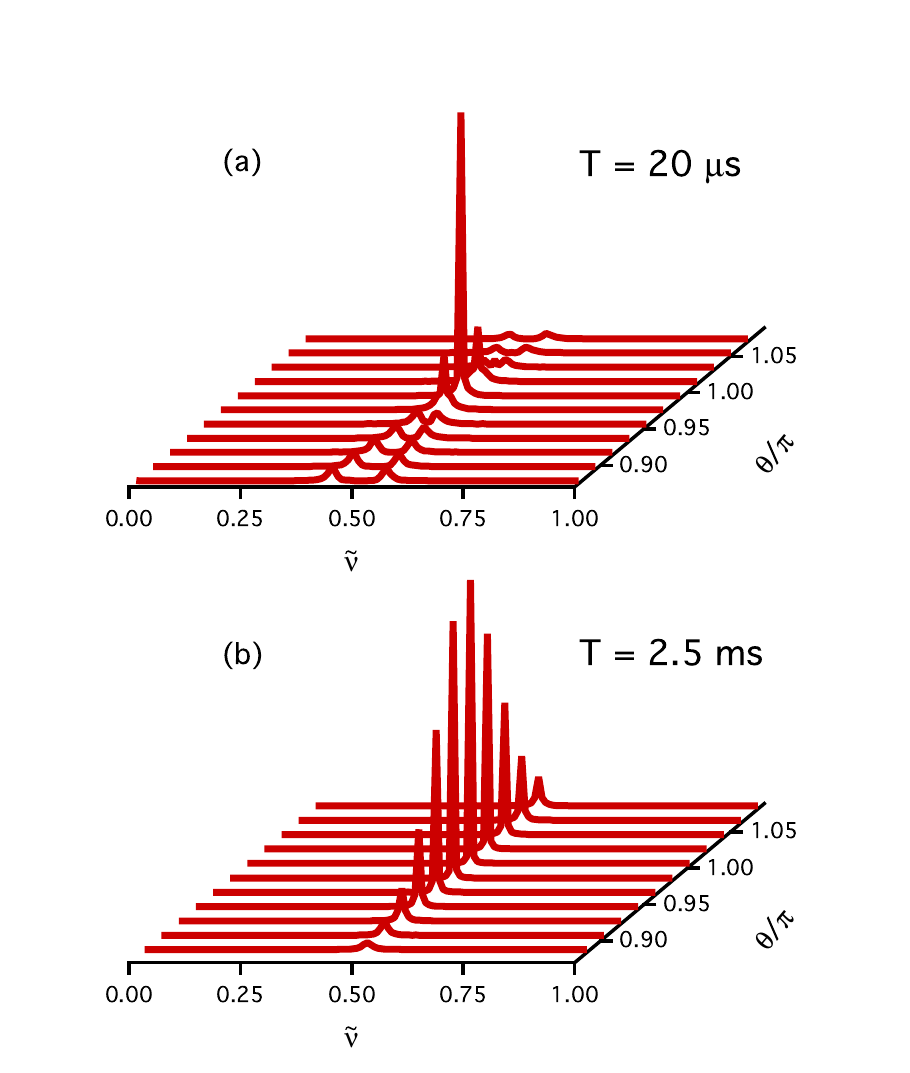}%
\caption{\label{WaterfallPlot}Waterfall plots showing the spectra $|S(\tilde{\nu})|^2$ at different $\theta$ with $^1$H off. Near $\theta=\pi$, a prominent subharmonic response is observed at $\tilde{\nu}=1/2$. (a) At short drive periods $T$, the subharmonic response splits into two peaks, which diverge almost immediately as $\theta$ is adjusted away from $\pi$. (b) For long drive periods $T$, the subharmonic peak lowers in amplitude as $\theta$ is adjusted away from $\pi$, but remains rigidly locked at $\tilde{\nu}=1/2$.}
\end{figure}

\section{EXPERIMENTAL SETUP AND DTC PULSE SEQUENCE}

The equilibrium $^{31}$P spins begin in a weakly polarized state described, as discussed above, by a reduced initial density matrix $\rho_0=I_{z_T}$. We improve the polarization and accelerate the experiments by instead exploiting the more highly polarized $^1$H spin bath as a source for CP. This provides a small improvement to the initial polarization of the $^{31}$P spins [$\rho_0'=(\gamma_\textrm{H} / \gamma_\textrm{P}) I_{z_T}$, $\gamma_\textrm{H} / \gamma_\textrm{P} \approx 2.5$], and a dramatic improvement to the repetition rate of the experiments since the $^1$H lattice relaxation time $T_1^\text{H} = 0.6$\,s is 200$\times$ faster than the $^{31}$P lattice relaxation time $T_1^\text{P}=103$\,s. To do this, we first excite the $^1$H spins with an initial $X_{\pi/2}$ pulse at the $^1$H frequency, followed by cross polarization with the $^{31}$P at the Hartmann-Hahn matching condition \cite{Hartmann1962, Pines1972, Pines1973}. This creates $^{31}$P $y$ polarization, which we convert to $z$ polarization with an $\overline{X}_{\pi/2}$ pulse at the phosphorus frequency (Fig. \ref{DTCSequence}). After each scan, we wait $3$\,s for the $^1$H to return to equilibrium (for the $T=1$\,s experiment described below, a $2$\,s wait time was used).

In order to look for evidence of discrete time translational symmetry breaking, we implement a ``DTC pulse sequence,'' consisting of a basic Floquet cycle which we repeatedly apply (Fig. \ref{DTCSequence}) following the preparation of the initial $I_{z_T}$ state. Each Floquet cycle is composed of a wait time $\tau$, during which the internal Hamiltonian is allowed to act freely, followed by a strong $X_{\theta}$ pulse of duration $t_p \approx 7.5$\,$\mu$s and angle $\theta = \pi + \epsilon$, with $|\epsilon/\pi| \ll 1$. This basic Floquet block is repeated $N$ times, represented as $\{\tau - X_{\theta}\}^N$. After $N$ cycles, we convert the $\braket{I_z}$ of the $^{31}$P spins into measurable transverse magnetization by applying a final, $X_{\pi/2}$ readout pulse. We measure the signal immediately after the pulse, which becomes the $N^\text{th}$ point in the data set [e.g., Fig. \ref{CompareToLukin}(a)]. Note that this is a slow incremental readout of the discrete time signal, since each repetition of the experiment allows us to choose only a particular value for $N$. Throughout this sequence we either allow the $^1$H to act on the phosphorus [``$^1$H on'', Fig. \ref{DTCSequence}(b)], or apply cw decoupling to the $^1$H, removing their effect on the $^{31}$P [``$^1$H off'', Fig. \ref{DTCSequence}(a)]. As we will discuss below, the rf power required for cw decoupling \footnote{For cw decoupling, we use $\gamma_\textrm{H} H_1 / 2 \pi \approx 18$\,kHz. This is a compromise between maximizing the decoupling performance and minimizing the heating of the NMR tank circuit.} will eventually heat (and detune) the circuit, limiting our ability to explore out to very long times with $^1$H off. 

To vary the applied pulse angle in our implementation of the DTC sequence, we maintain a constant $t_p$ and vary the strength of the pulse $\omega_1$. This gives us better resolution in the pulse angle, while maintaining a constant cycle period $T$. The internal Hamiltonian continues to act during a pulse, an important fact despite the short pulse duration --- we return to this in Sec. \ref{CausesOfDecay}. 

The DTC pulse sequence for a given $\theta$ and $\tau$ produces a discrete time signal $S(t)$ with $t=NT$, where the period of the two-step drive is $T=\tau + t_p$, and the corresponding frequency of the Floquet drive is $\nu_F=1/T$. We Fourier transform $S(t)$ to get the complex spectrum $S(\nu)$, then examine $|S(\tilde{\nu})|^2$ as a function of the normalized frequency $\tilde{\nu}=\nu / \nu_F$. For $N=1,2,...,128$, the normalized frequency $\tilde{\nu}$ takes discrete values from $0$ to $127/128$, in steps of $d\tilde{\nu}=1/128$.

 \section{RESULTS: DTC OSCILLATIONS OVER A RANGE OF \textbf{$\theta$} AND MANY DECADES OF \textbf{$\tau$}}
 
 First, we discuss noteworthy features of the $^1$H-off data set. When we apply the DTC pulse sequence for $\theta \approx \pi$ ($|\epsilon/\pi| < 0.01$) and at small time $T=20$\,$\mu$s, $S(t)$ follows what intuition would dictate, trivially reversing its sign with each successive Floquet cycle. This corresponds to a single Fourier peak at normalized frequency $\tilde{\nu}=1/2$ [Fig. \ref{CompareToLukin}(a)]. When $\theta$ is adjusted away from $\pi$, still at small $T=20$\,$\mu$s, there is a pronounced additional modulation of the signal, corresponding to a splitting of the Fourier peak --- again an expected result [Fig. \ref{CompareToLukin}(b)]. However, at the same approximate deviation $|\epsilon/\pi| = 0.04$, if we increase $\tau$ such that $T = 400$\,$\mu$s (giving the dipolar interaction a longer time to act), the single Fourier peak at $\tilde{\nu}=1/2$ is restored [Fig. \ref{CompareToLukin}(c)]. Figure \ref{CompareToLukin}(c) shows a predicted signature of the DTC \cite{Yao2017}: at long enough $\tau$, the oscillations in $S(t)$ are rigidly locked at $\tilde{\nu}=1/2$, despite adjusting $\theta$ away from $\pi$. Increasing $T$ by more than an order of magnitude, we are still able to observe the locked oscillations [Fig. \ref{CompareToLukin}(d)]. For brevity, we refer to $S(t)$ signals such as those in Fig. \ref{CompareToLukin}(c) as ``DTC oscillations.'' When we conduct a comparable experiment but with $^1$H on, we observe very similar behavior [Figs. \ref{CompareToLukin}(e)-\ref{CompareToLukin}(h)].

When we apply the DTC pulse sequence for many values of $\theta$ at $T=20$\,$\mu$s, we observe that the prominent feature at $\tilde{\nu}=1/2$ splits into two separate frequencies, which grow apart as $\theta$ deviates from $\pi$ [Fig. \ref{WaterfallPlot}(a)]. However, for a longer drive period $T=2.5$\,ms, this fails to happen. The response at $\tilde{\nu}=1/2$ instead remains locked in place, while diminishing in height as $\theta$ deviates from $\pi$ [Fig. \ref{WaterfallPlot}(b)].

\begin{figure}
  \includegraphics[width=0.45\textwidth]{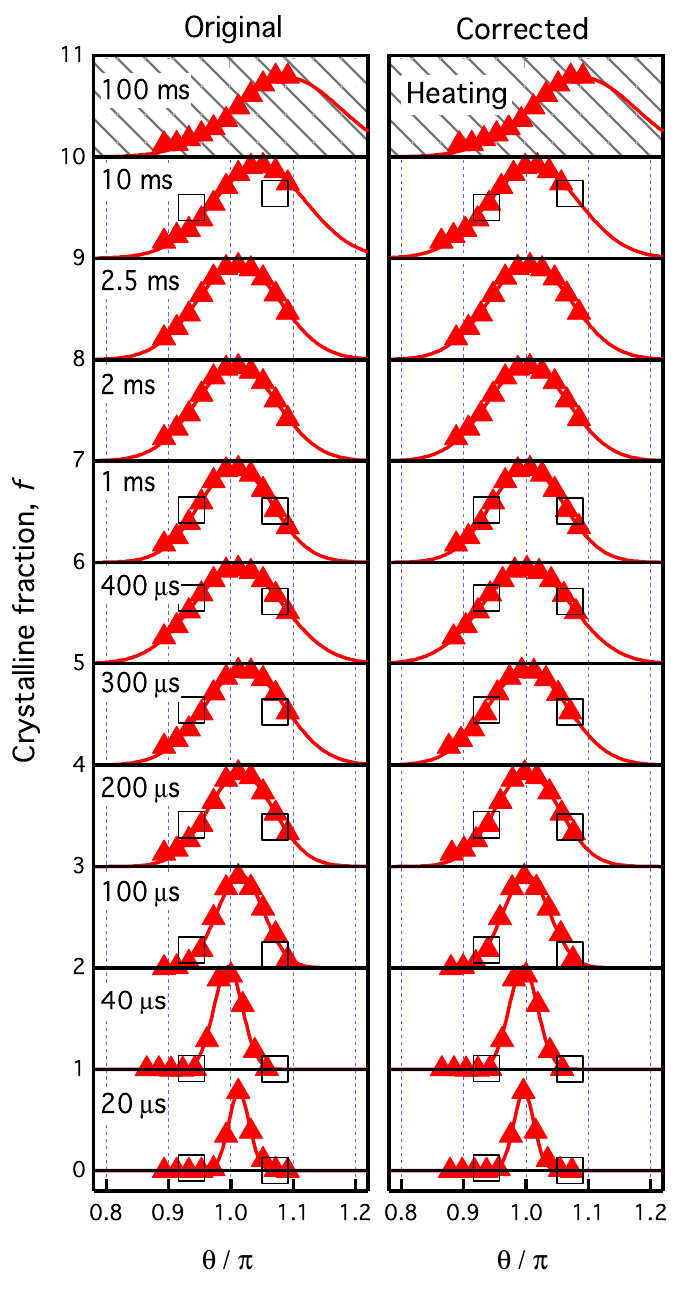}%
\caption{\label{GaussiansCW}$^1$H off: crystalline fractions $f$ across a broad range of drive periods $T$ (labeled, where the $f$ data for the $m$th value of $T$ are vertically offset by $m-1$ for clarity). The crystalline fractions are well fit by Gaussians. Over the duration of the many experiments, the tuning of the NMR tank circuit can drift, leading to poorly calibrated $\theta$ (left). The black squares represent well-calibrated benchmarks which we use to correct the data to match the actual $\theta$ values (as described in the main text), resulting in the data on the right. Because of heating, the $100$-ms data will not appear in Fig. \ref{PhaseBound}. Error bars (not shown) are much smaller than the markers.}
\end{figure}

\begin{figure}
  \includegraphics[width=0.45\textwidth]{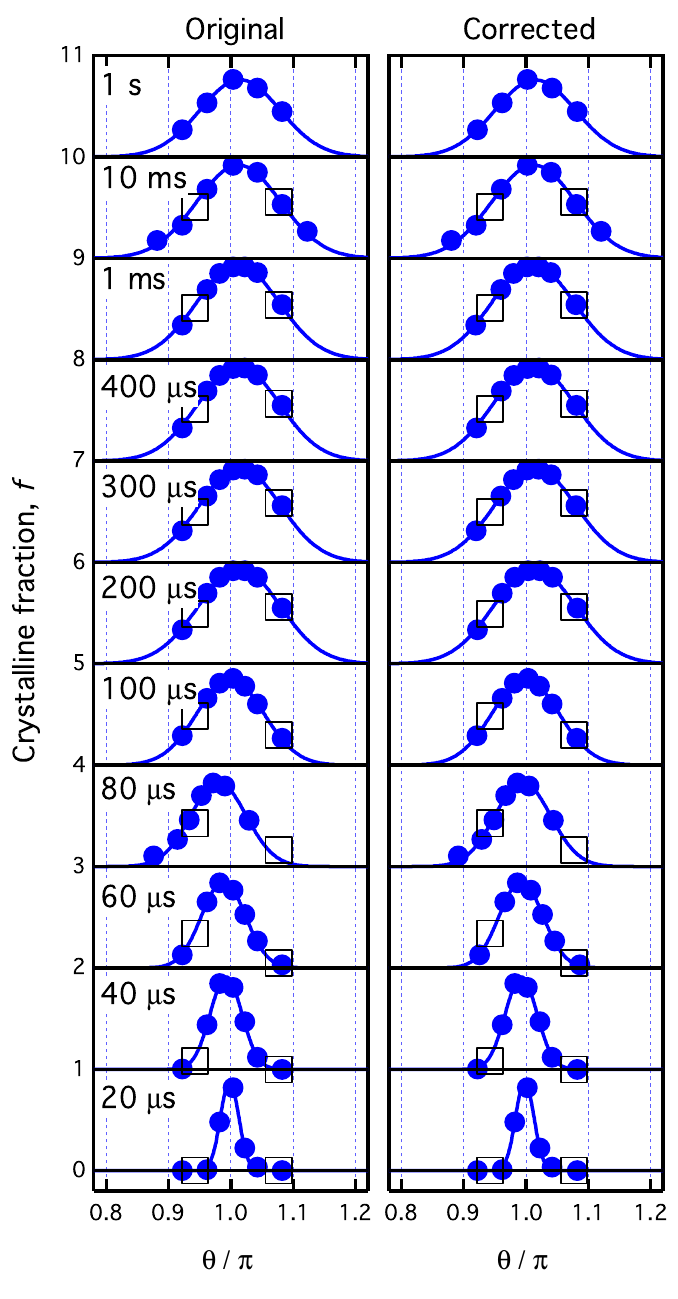}%
\caption{\label{GaussiansNoCW}$^1$H on: In the absence of $^1$H decoupling, we are free to explore an even greater expanse in $T$ (labeled, where the $f$ data for the $m$th value of $T$ are vertically offset by $m-1$ for clarity) without the danger of the circuit heating effects seen in Fig. \ref{GaussiansCW}. As with $^1$H off, we observe Gaussian shapes in the crystalline fraction with $^1$H on, and the width of the Gaussians increases with the drive period. The black squares are used in the same correction procedure as those shown in Fig. \ref{GaussiansCW}, to correct for miscalibrations of the actual $\theta$ from the expected $\theta$ (a very minor effect here). Error bars (not shown) are much smaller than the markers.}
\end{figure}

To characterize the response to the DTC pulse sequence across the $(\theta,\tau)$ plane, we examine the ``crystalline fraction'' as introduced by Choi \textit{et al}.\ \cite{Choi2017}: $f=|S(\tilde{\nu}{=}1/2)|^2/\sum_{\tilde{\nu}}|S(\tilde{\nu})|^2$. For each value of $\tau$, we vary $\theta$ around $\pi$ (by varying $\omega_1$ at fixed $t_p$) and plot the crystalline fractions, using all 128 points of $S(t)$, which fit well to Gaussians (Figs. \ref{GaussiansCW} and \ref{GaussiansNoCW}). A set of crystalline fraction measurements at a single $\tau$ typically takes about one day to complete; over the course of many such experiments, the tuning of the NMR tank circuit may drift, leading to slight changes in the actual $\theta$ compared to the intended $\theta$. To correct for this, we recalibrated $\omega_1 t_p = \theta$ (using a nutation experiment), then conducted two experiments at constant $\theta$ and varying $\tau$, allowing us to explore along $\tau$ relatively quickly after the calibration (Figs. \ref{GaussiansCW} and \ref{GaussiansNoCW}, black squares). By running two such experiments, we are able to use the resulting crystalline fractions as ``guides'' to line up the data across experiments, correcting for the slow drift in pulse power and reducing systematic uncertainty in $\theta / \pi$ (Figs. \ref{GaussiansCW},\ref{GaussiansNoCW}).

\begin{figure}
\includegraphics[width=0.45\textwidth]{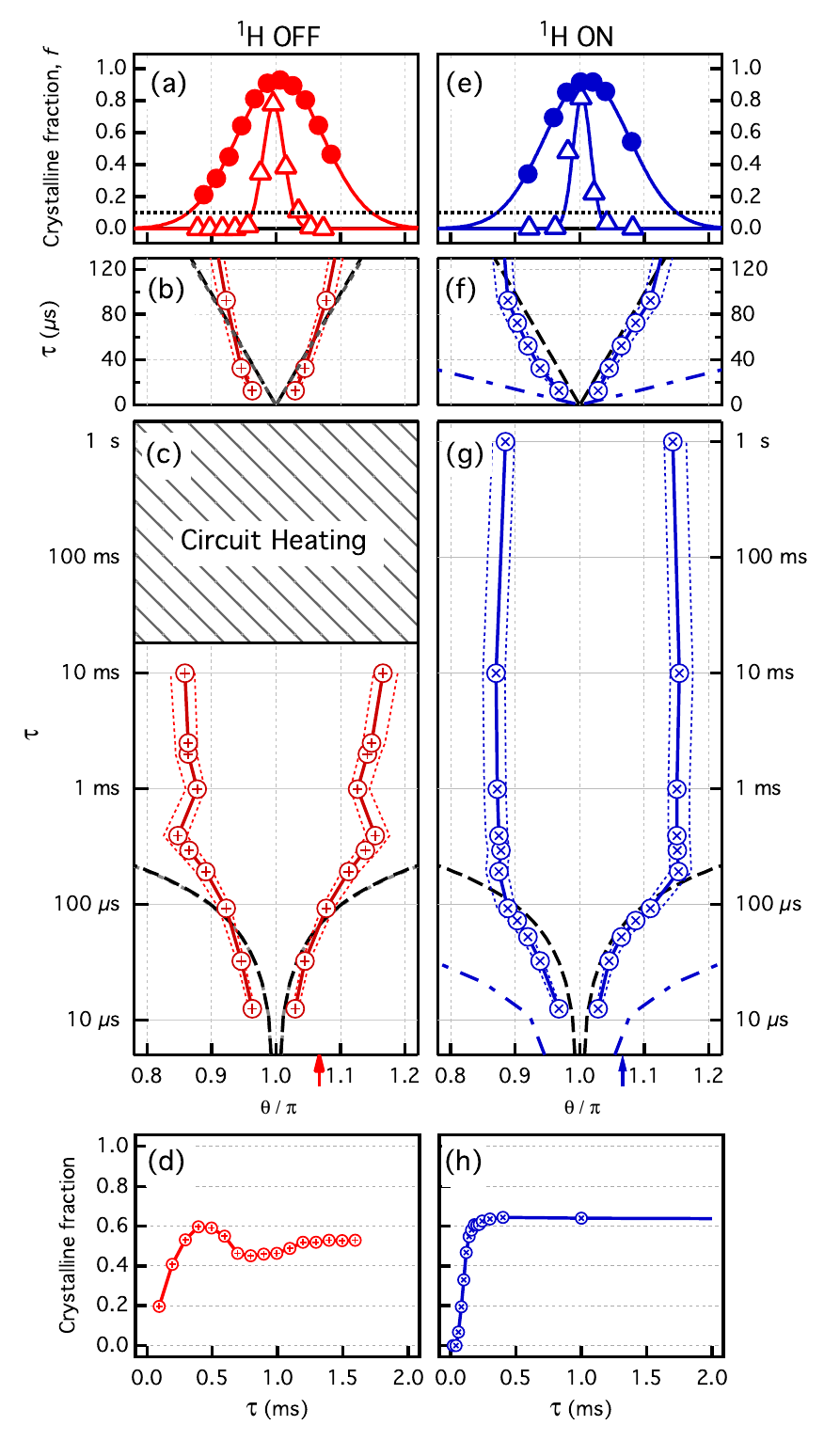}%
\caption{\label{PhaseBound}(a)-(d) Probing $f$ with $^1$H off. (a) We establish a cutoff (dotted line) in the Gaussian fits to the crystalline fraction at $f=0.1$. Crystalline fractions from $T=20$\,$\mu$s (triangles) and $400$\,$\mu$s (circles) are shown. The intersection of $f$ with the cutoff defines a boundary point. (b) Cutoff at $f=0.1$ (red circles), corresponding to the boundaries within which we observe persistent oscillations at $\tilde{\nu}=1/2$ (the ``DTC region''). We also show cutoffs at $f=0.05$ and $f=0.15$ (dotted lines). We compare this to an effective dipolar interaction angle by plotting $|\theta - \pi| = W \tau$, with $W=W^\text{P,P}$ [(b-c), black dashed lines] and $W=W^\text{P,PN}$ [(b-c), gray dashed lines, very close to $W^\text{P,P}$]. (c) DTC region on a semi-log scale. For $\tau=100$\,ms, the results in Fig. \ref{GaussiansCW} become unreliable because of tank circuit heating from rf decoupling, so they are not plotted here. (d) $f$ versus $\tau$ for $^1$H off at $\theta=1.067\pi$ [angle marked in (c)]. (e-h) Probing $f$ with $^1$H on. In (f-g) we also include $|\theta - \pi| = W^\text{P,HPN} \tau$ (blue dotted-dashed lines). In (g), the data span the range $0.03 < W^\text{P,P}\tau < 3200$\,radians. Error bars (not shown) are much smaller than the markers in (a)-(h).}
\end{figure}

Following the example of Choi \textit{et al}.\ \cite{Choi2017} once again, we visualize the region of persistent DTC oscillations by noting where the Gaussian fits to the crystalline fractions fall below an arbitrary cutoff [Figs. \ref{PhaseBound}(a) and \ref{PhaseBound}(e)]. We show the corresponding $(\theta,\tau)$ values for crystalline fraction $f=0.1$, along with those for $f=0.05$ and $0.15$, since the region exhibiting persistent DTC oscillations does not show particularly sharp boundaries. The resulting diagrams shown in Figs. \ref{PhaseBound}(b) and \ref{PhaseBound}(c) and \ref{PhaseBound}(f) and \ref{PhaseBound}(g) depict the boundaries within which we observe DTC oscillations (the ``DTC region''), and outside of which the deviation of the drive from $\theta=\pi$ results in diminished or split Fourier peaks in the spectrum. At small $\tau$, there exists a very small region of DTC oscillations around $\theta=\pi$. As $\tau$ is increased, the oscillations persist for a wider and wider range of $\theta$ around $\pi$, as the DTC region ``expands'' in width. For both $^1$H on and $^1$H off, at long $\tau$, the width of the DTC region becomes roughly independent of $\tau$ over multiple orders of magnitude. We do not observe a predicted ``pinch-off'' of the stable region at long $\tau$, perhaps because our spin Hamiltonian does not have the disorder assumed in that model \cite{Ho2017}. For $^1$H off [Fig. \ref{PhaseBound}(c)], we observe some structure in the the DTC boundary around $\tau=1$\,ms. For $^1$H on [Fig. \ref{PhaseBound}(g)], the width of the DTC region increases slightly faster at short $\tau$, and is relatively featureless at long $\tau$ compared to the $^1$H off case.

When the rf power from $^1$H decoupling causes circuit heating, there can be different amounts of heating at different $N$ values. This makes it very difficult to calibrate the results, so we omit the data acquired in the presence of significant circuit heating (Fig. \ref{GaussiansCW}, $T=100$ ms) from Fig. \ref{PhaseBound}(c). When we repeat these experiments with $^1$H on (with no cw decoupling and no circuit heating), we are able to explore even more decades in $T$ [Figs. \ref{GaussiansNoCW} and \ref{PhaseBound}(g)], out to $T=1$\,s, where the total experiment time approaches $T_1^\text{P}$. This is likely responsible for the slight decrease in the crystalline fraction amplitude at $T=1$\,s, and the corresponding decrease in the width of the DTC region at $T=1$\,s (note that the Gaussian fit in Fig. \ref{GaussiansNoCW}, $T=1$\,s, is shorter than those at smaller $T$, rather than narrower).

\begin{figure}
\includegraphics[width=0.45\textwidth]{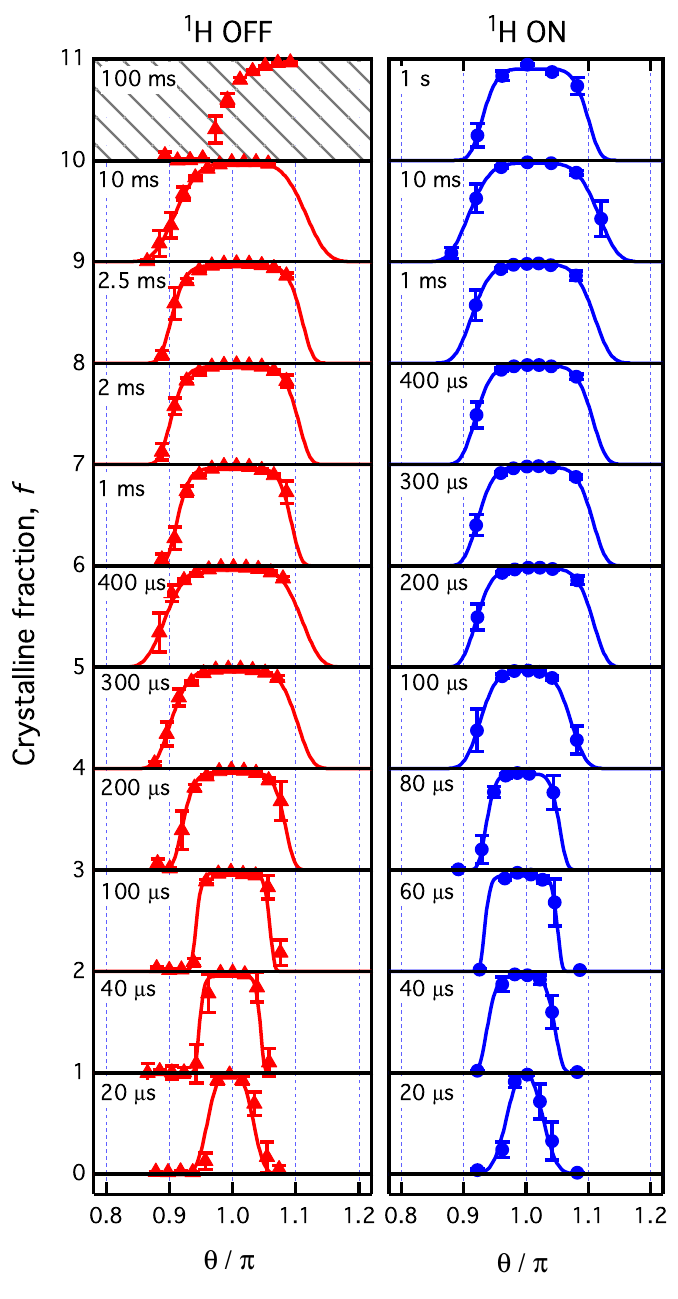}%
\caption{\label{SuperGaussians}Using Fourier transforms of only 50 late-time points in $S(NT)$, $N=51$---$100$, the crystalline fractions become flatter around $\theta = \pi$, for both $^1$H off (left, red triangles) and $^1$H on (right, blue circles). We fit these to symmetrical super-Gaussians (lines): $F(\theta)=A\exp[-(|\theta-\theta_0|/\sigma)^p/2]$, where we fix $\theta_0$ using the Gaussians in Figs. \ref{GaussiansCW} and \ref{GaussiansNoCW}.}
\end{figure}

To provide a unitless scale for these results, we compare the deviations of the rf pulse angle $\theta$ from $\pi$ to an effective dipolar interaction angle $W\tau$. We show lines at $|\theta-\pi| = W\tau$, for $W^\text{P,P}/2\pi= 508$ Hz [Figs. \ref{PhaseBound}(b), \ref{PhaseBound}(c), \ref{PhaseBound}(f), \ref{PhaseBound}(g)], $W^\text{P,PN}/2\pi= 517$ Hz [Figs. \ref{PhaseBound}(b) and \ref{PhaseBound}(c)], and $W^\text{P,HPN}/2\pi = 3538$\,Hz [Figs. \ref{PhaseBound}(f) and \ref{PhaseBound}(g)]. These lines are not considered to be explanations for the shape of the DTC boundary, but it is interesting that they are so close to the boundary at small $\tau$. To better understand the non-monotonic, complicated structure in the boundary of the DTC region around $\tau = 1 $\,ms for $^1$H off [Fig. \ref{PhaseBound}(c)], we reexamine the crystalline fraction with an experiment at fixed $\theta = 1.067\pi$ and linear scales in $\tau$ for both $^1$H on and $^1$H off [Fig. \ref{PhaseBound}(d,h), which show crystalline fractions rather than cutoff boundaries]. In Fig. \ref{PhaseBound}(d), we see the crystalline fraction is a non-monotonic function of $\tau$ for $^1$H off. By contrast, Fig. \ref{PhaseBound}(h) shows that the crystalline fraction for $^1$H on has a steeper slope at short $\tau$, and is without structure at long $\tau$.

The Gaussian shapes shown in Figs. \ref{GaussiansCW} and \ref{GaussiansNoCW} differ from the corresponding super-Gaussian shapes reported by Choi \textit{et al}. [their Fig. 3(a)] \cite{Choi2017}. While this might seem to be an important difference, it turns out to be an artifact of the FT window size used in each study. To see this, we recalculate our crystalline fractions using a windowed FT of only the points $N=51$--$100$ in $S(t)$, which matches exactly the procedure of Choi \textit{et al}. \cite{Choi2017}. Figure \ref{SuperGaussians} shows that the resulting data are much flatter near $\theta=\pi$, and are well described by Choi \textit{et al}.'s super-Gaussian model. At first glance, the impact of window-size choice on crystalline-fraction shape seems paradoxical, since our $S(t)$ data are typically single exponential [e.g., Fig. \ref{CompareToLukin}(d)]. However, the crystalline fraction should not be confused with a time constant, as its value depends on the choice of the FT window in a complicated way (for a more in-depth explanation, see Appendix \ref{WindowDependence}). In light of this, we think it is best to use our full data sets when calculating the crystalline fraction. 

It is interesting to note the similarity between the results for $^1$H off and $^1$H on, as well as the similarity to the results achieved using diamond NV centers, despite the different spin Hamiltonians. Note also that the $W \tau$ range shown is $0.03 < W^\text{P,P}\tau < 3200$\,radians, spanning effective dipolar interaction angles both far below and far above $W\tau=1$\,radian. In most DTC models, thermalization should destroy the oscillating signal for long enough $\tau$, but we do not see this in our results.

  \section{\label{Refocusing}REFOCUSING THE DECAY OF THE DTC OSCILLATIONS WITH THE DTC ECHO SEQUENCE}

  The lifetime of the DTC oscillations (and the dependence of lifetime on the interaction strength) is of central interest in the study of DTC physics. To explain the observed decay in our experiments, we first consider a simple model of noninteracting spins, which undergo a two-step process starting with magnetization along $\hat{z}$. First, an $X_{\pi + \epsilon}$ pulse rotates the magnetization vector to $-\hat{z}\cos(\epsilon) - \hat{y}\sin(\epsilon)$. Second, during the time $\tau$, we suppose that the transverse magnetization is lost due to dephasing caused by local field variations, leaving only the component of the magnetization along $\hat{z}$. After repeating this process over $N$ cycles, the original signal will have decayed exponentially as $\cos^N(\epsilon)$. Indeed, the signal we observe in our experiments seems to stay at or below the bound imposed by this predicted decay envelope. If the dephasing in our model is due to external field variations of unknown origin, then this decay will be irreversible. On the other hand, if the observed decay is actually due to unitary evolution under a complicated Hamiltonian, then it might, in principle, be reversible. To test whether this decay was reversible or not, we devised a pulse sequence designed to undo the forward evolution from the dominant Hamiltonian terms, looking for instances where the signal rose above the envelope imposed by the $\cos^N(\epsilon)$ decay model.

 \begin{figure}
   \includegraphics[width=0.45\textwidth]{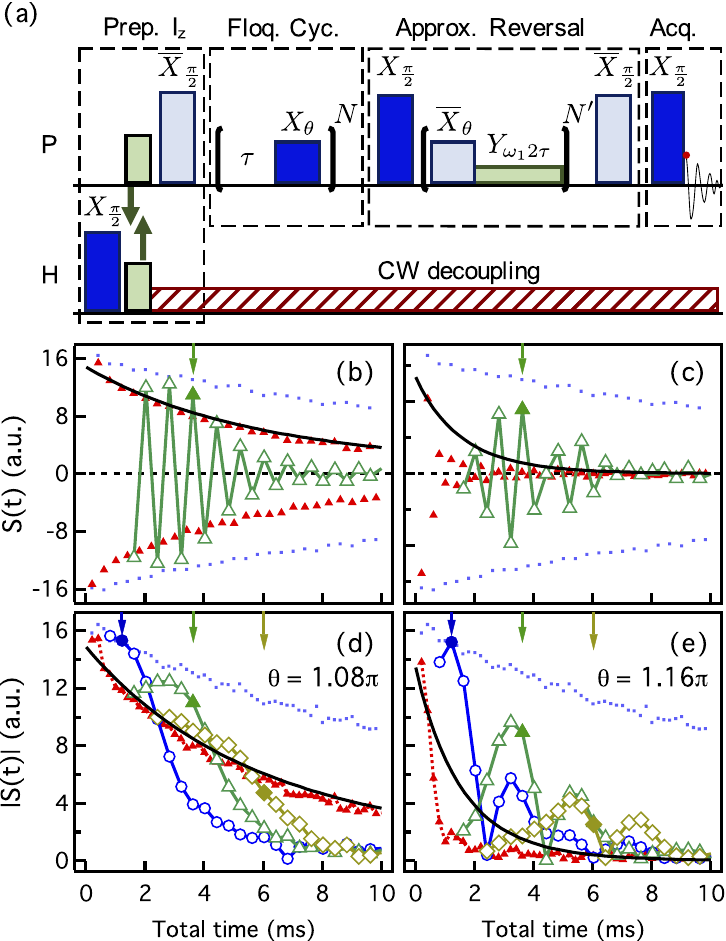}%
\caption{\label{Echoes}(a) DTC echo sequence, designed to approximately reverse the effect of the original DTC sequence. The ``approximate reversal'' block consists of a rotation $\overline{X}_\theta$ (wide orange block), followed by a duration $2\tau$ during which a strong pulse of phase $y$ is applied to the $^{31}$P. We apply ``wrapper'' pulses $X_{\pi/2}$ and $\overline{X}_{\pi/2}$ (tall blue and orange blocks, respectively) to rotate $-\mathcal{H}^\text{P,P}_{yy}$ into $-\mathcal{H}^\text{P,P}_{zz}$. Since the last two pulses of the sequence negate one another, neither is applied in practice. $^1$H decoupling is used throughout. (b), (c) DTC echoes for $T=200$\,$\mu$s and $\theta = 1.08\pi$ (b) and $1.16\pi$ (c). For $N$ cycles of the ``forward'' block, we see the signal decay in red closed triangles. After $N=6$, the reversal sequence is applied for $N'$ cycles (green open triangles), where we expect an echo to appear at $N' = N = 6$ (filled point and arrow). (d), (e) DTC echoes for $N=\{2,6,10\}$ (open blue circles, green triangles, yellow diamonds), where we show the absolute values of each signal for easier inspection. Expected echo locations are marked with filled points and arrows. In (b)-(e), blue dots show the DTC signal decay for $\theta \approx \pi$.}
\end{figure}

If we assume the effect of $\mathcal{H}_\text{int}$ during $\tau$ is dominated by the $^{31}$P-$^{31}$P dipolar coupling $\mathcal{H}_{zz}^\text{P,P}$, then we can borrow techniques from the ``magic-echo'' experiment, which is designed to refocus the homonuclear dipolar interaction \cite{Rhim1971}. To adapt these techniques for designing a ``DTC echo'' sequence, we use two approximations. First, we assume that all of the short duration ($<10$\,$\mu$s) applied $X_{\theta}$ pulses are of infinite strength and zero duration, such that the net rotation angle is $\theta$ and the internal Hamiltonian has no time to act (i.e., the delta-function pulse approximation). Second, during a much longer pulse ($\gg 10$\,$\mu$s) of phase $\phi$, we assume that the homonuclear dipolar coupling reduces exactly to the component of the dipolar coupling which is secular in the frame of the pulse: $\mathcal{H}^\text{P,P}_{zz}\rightarrow -(1/2)\mathcal{H}^\text{P,P}_{\phi \phi}$, where we have defined $\mathcal{H}^\text{P,P}_{\phi \phi}~=~\sum_{i,j>i}B^P_{ij} ( 3I_{\phi_i}I_{\phi_j}~-~\vec{I}_i~\cdot~\vec{I}_j )$ \cite{Slichter1996}. Using these approximations, we construct a unitary reversal of the original DTC Floquet cycle by time-reversing both the $X_{\theta}$ pulse and the free evolution, in reverse order. To reverse the effect of the $X_{\theta}$ pulse, we simply apply a pulse of equal angle but opposite phase, $\overline{X}_{\theta}$. To reverse the effect of the homonuclear dipolar term in the internal Hamiltonian, $\mathcal{H}_{zz}^\text{P,P}$, we make use of the abovementioned approximation, and apply a long $Y_\Phi$ pulse, where $\Phi=\omega_1 2 \tau$, to produce an effective evolution of $(-1/2)\mathcal{H}^\text{P,P}_{yy}(2\tau) = -\mathcal{H}^\text{P,P}_{yy}\tau$. In order to properly negate the forward evolution $\mathcal{H}^\text{P,P}_{zz}\tau$ from the original sequence, we include ``wrapper pulses'' $\pm X_{\pi/2}$ around $-\mathcal{H}^\text{P,P}_{yy}\tau$, which ``rotate'' it into $-\mathcal{H}^\text{P,P}_{zz}\tau$. The resulting DTC echo sequence is:
 \begin{equation}
\{\tau - X_\theta\}^N- (X_{\pi/2} - \{ \overline{X}_\theta - Y_{\Phi} \}^{N'} - \overline{X}_{\pi/2}).
 \end{equation}
 This is shown schematically in Fig. \ref{Echoes}(a). Starting with $N$ Floquet cycles of the original DTC pulse sequence, we follow with $N'$ repetitions of the approximate reversal sequence, looking for an echo peak when $N'=N$. In the language of the more conventional Hahn spin echo sequence, the first part ($N$ blocks) of this sequence generates the ``FID'' analog, while the second part (rotated $N'$ blocks) generates the echo signal ``after the $\pi$ pulse'' \cite{Hahn1950,Slichter1996}. Note that this DTC echo sequence would not be able to refocus the decay of the DTC oscillations if it were instead dominated by a spread in static Zeeman terms $\Omega_i I_{z_i}$, because the strong $Y_{\Phi}$ pulse quickly averages these Zeeman terms to zero during the $N'$ blocks of the DTC echo sequence.

 Using the original DTC pulse sequence for $\theta=1.08\pi$ [Fig. \ref{Echoes}(b)] and $\theta=1.16\pi$ [Fig. \ref{Echoes}(c)], we see $S(t)$ decay near or below the $\cos^N(\epsilon)$ decay rate. Using the DTC echo sequence for $N=6$ Floquet cycles of the original DTC sequence, we observe clear echoes rising above the $\cos^N(\epsilon)$ decay envelope [Figs. \ref{Echoes}(b) and \ref{Echoes}(c)]; these echoes are even more prominent when we plot $|S(t)|$ for the DTC echo sequence with $N=2,6,10$ [Figs. \ref{Echoes}(d) and \ref{Echoes}(e)]. This demonstrates that the decay mechanism of the DTC oscillations involves deterministic coherence flow to unobservable parts of the density matrix, which our DTC echo sequence then resurrects as signal.
 
\begin{figure}
\includegraphics[width=0.5\textwidth]{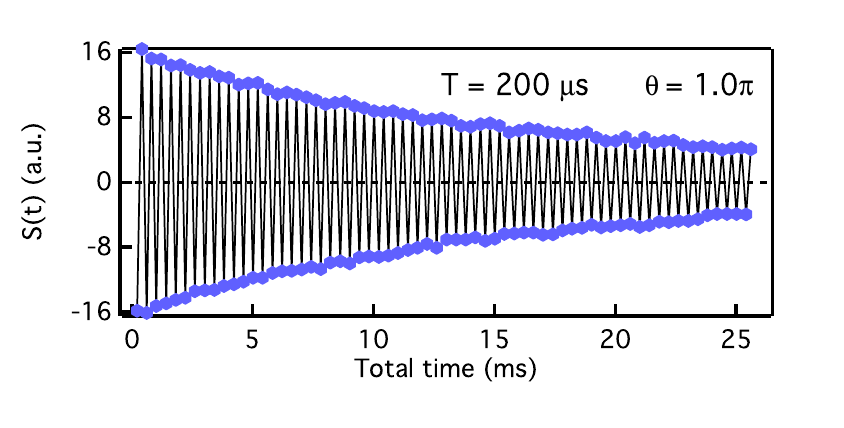}%
\caption{\label{DecayAtPi}$S(t)$ for $\theta = \pi$, at $T=200$\,$\mu$s, identical to the blue dots in Fig. \ref{Echoes}(b-e). This decay cannot be explained by dipolar interactions in an ideal delta-function pulse model.}
\end{figure}

  \section{\label{CausesOfDecay}CAUSES OF THE DECAY IN THE DTC OSCILLATIONS OBSERVED AT \textbf{$\theta=\pi$}}

  For our spin Hamiltonian, we do not expect to see any decay in the DTC oscillations at $\theta=\pi$, if we apply perfect, delta-function pulses. However, Fig. \ref{DecayAtPi} shows that the oscillations clearly decay even at $\theta=\pi$, which causes us to revisit the effects of the actual pulses used in the DTC sequence. Another clue about the mechanism responsible for this decay is that it seems to impose a limit on the echoes produced in the $\theta > \pi$ case (Fig. \ref{Echoes}), where the echoes never rise above the data acquired at $\theta = \pi$ (Fig. \ref{Echoes}, blue dots), and also appear to occur slightly earlier than expected as if there is an additional decay envelope imposed on their evolution. In this section we discuss possible causes for this decay envelope, first by examining possible experimental causes, then by revisiting the approximation of zero-duration, delta-function $\pi$ pulses.

  \subsection{Quantifying the effect of experimental pulse imperfections}

 \begin{figure}
\includegraphics[width=0.5\textwidth]{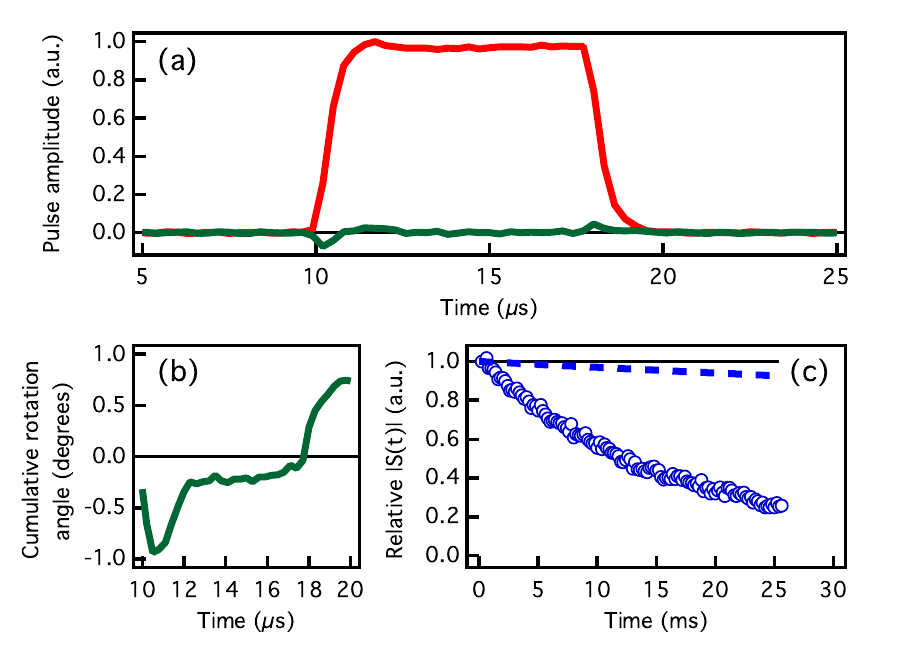}%
\caption{\label{PhaseTransients}(a) Envelope of a $\pi$ pulse applied at frequency $\omega_0$, as measured by a pickup coil placed near the resonator. The phase transients (small out-of-phase signal, dark green) are much smaller than the in-phase (red) pulse amplitude. (b) Cumulative integral of the out-of-phase pulse amplitude, scaled by setting the integral of the in-phase signal to 180$\degree$. Each transient produces less than a degree of rotation. (c) Incorporating the effect of phase transients into the $\cos^N(\epsilon)$ decay model (black, identically 1 for $\theta=\pi$) leads to a modified decay model (dashed line). The modified model decays too slowly to account for the decay envelope of the measured $|S(t)|$ at $\theta=\pi$, shown here for $T=200$\,$\mu$s (blue circles).}
\end{figure}

  In this section, we consider two common pulse imperfections, phase transients and $H_1$ inhomogeneity, and quantify their effects on the decay envelope at $\theta=\pi$. First we consider phase transients, which are small out-of-phase components of the applied rf at the beginning and end of the pulses. Using a small pickup coil connected to the NMR spectrometer acquisition channel, we measure the applied magnetic field from a pulse, and compare the out-of-phase component to the in-phase $\pi$ pulse. As evident in Fig. \ref{PhaseTransients}(a), the measured phase transients are very small relative to the in-phase component of the applied $X_\pi$ pulse. The cumulative effect of these transients results in a net out-of-phase (along $\hat{y}$) rotation of less than $1^{\circ}$ [Fig. \ref{PhaseTransients}(b)]. To incorporate this into the ``product-of-cosines'' decay model, we model a pulse with phase transients by including small out-of-phase components of opposite sign before and after the intended pulse: $X_\theta \rightarrow \{Y_{1^{\circ}}-X_\theta-Y_{-1^\circ}\}$ \cite{Li2008}. Then we again assume that after each such pulse, only the magnetization along $\hat{z}$ remains. This modifies the original product-of-cosines model to $\cos^N(\epsilon) \rightarrow [\cos^2(1^\circ)\cos(\epsilon) - \sin^2(1^\circ)]^N$, whose magnitude we show in Fig. \ref{PhaseTransients}(c). Comparing to the DTC oscillations at $\theta=\pi$, we see that the effect of the phase transients is far too small to account for the observed decay envelope.

  Second, we consider the effect of $H_1$ inhomogeneity across the sample, due to the coil geometry. If $H_1$ varies across the sample, then an applied pulse of intended angle $\theta$ will actually produce rotations of slightly different angles in different parts of the sample. To investigate this, we carry out a nutation experiment, which examines the signal after a pulse $X_{\omega_1t}$, where $\omega_1$ is constant and the pulse time $t$ is stepped from small to large values. Fitting the signal as a function of pulse time $t$ reveals the frequency of oscillation, $\omega_1$. The envelope of this long nutation curve will decay for two reasons: the $H_1$ inhomogeneity across the sample, and the reduced homonuclear dipolar coupling $(-1/2)\mathcal{H}_{xx}^\text{P,P}$ during the long pulse. In order to arrive at a decay model which incorporates the $H_1$ inhomogeneity, we do three things. First, we quantify the decay caused by $(-1/2)\mathcal{H}_{xx}^\text{P,P}$, and ``remove'' this effect from the nutation curve, leaving only the decay due to the $H_1$ spread across the sample. Second, we use this altered nutation data to infer a probability distribution of $H_1$ strength across the sample. Third, we use this inferred distribution to create a modified ``product of cosines'' decay envelope which takes into account the spread in the applied angle. The details of each step follow.

\begin{figure}
\includegraphics[width=0.5\textwidth]{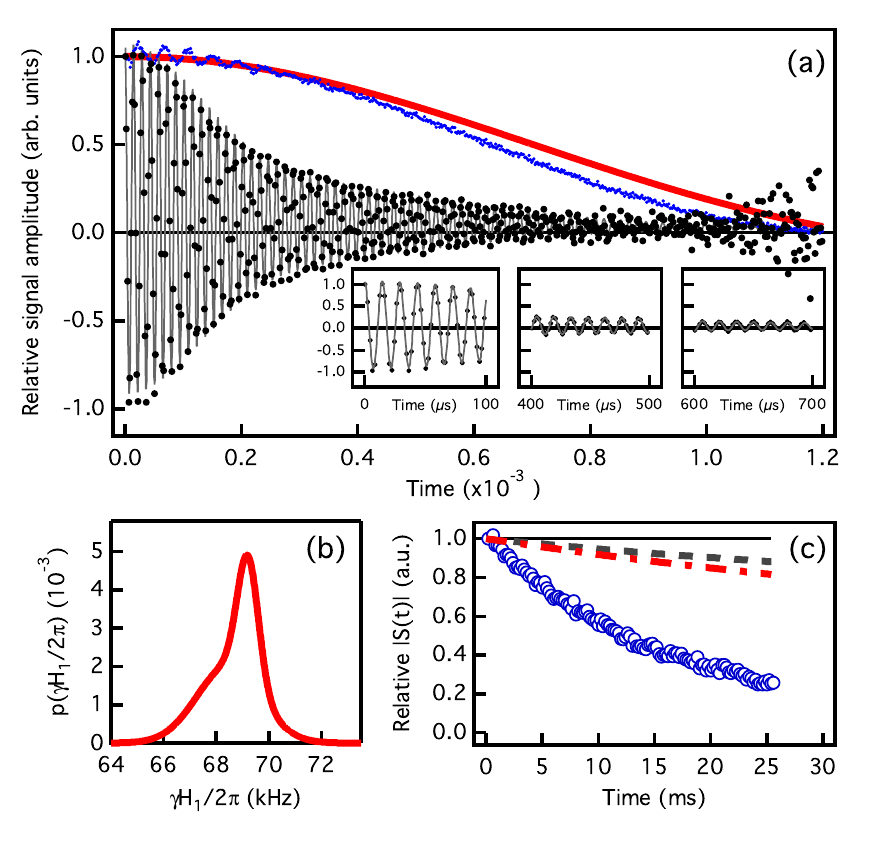}%
\caption{\label{NutationRotaryEcho}(a) Nutation of $^{31}$P (black dots), after removing the effect of the homonuclear dipolar coupling. We do this by dividing the measured nutation data (not shown) by the simulated Hahn echo, scaled by a factor of 2 in time (red solid line). A rotary echo experiment approximately undoes the effect of the $H_1$ inhomogeneity, and results in data (small blue dots) which match closely to the scaled Hahn echo simulation. We fit (gray) the modified nutation data to a model of a plausible field profile in the coil, with good results (a, inset). (b) The $H_1$ field profile corresponding to the fit parameters (from the fit to the scaled nutation data) for a two-Gaussian model; this represents a histogram of applied frequencies $\gamma H_1 / 2\pi$ during an applied pulse. (c) The histogram of applied frequencies $\gamma H_1/2\pi$ can be used to deduce a spread in the applied pulse angle $\theta$, which we use to modify the original product-of-cosines decay model (black, identically 1 for perfect $\pi$ pulses) to the corrected decay model (gray dashed line). Including the effects of both $H_1$ inhomogeneity and phase transients (red dotted-dashed line) only slightly modifies the modeled decay envelope. The magnitude of $S(t)$ at $\theta \approx \pi$ and $T=200$\,$\mu$s (blue circles) decays much faster than the product-of-cosines model, even after including the effect of these pulse imperfections.}
\end{figure}

 We quantify the decay caused by $(-1/2)\mathcal{H}_{xx}^\text{P,P}$ using both a rotary echo experiment and a Hahn echo experiment. In a rotary echo experiment \cite{Solomon1959}, a data point at time $t$ is acquired by examining the signal after applying the pulse sequence $\{X_{\omega_1t/2}-\overline{X}_{\omega_1t/2}\}$ at constant $\omega_1$ (note that we require 2\,$\mu$s gaps between consecutive pulses to change phase, which we do not show in the pulse sequence notation here or below). This approximately negates the spread from the applied field inhomogeneity, leaving only the decay due to the component of dipolar coupling that is secular in the presence of the strong $\pm \hat{x}$ pulse: $-(1/2)\mathcal{H}^\text{P,P}_{xx}$. Since this reduced dipolar coupling has a prefactor of $1/2$, we expect a rotary echo experiment to produce similar results to those of a Hahn echo experiment with time effectively doubled. In Fig. \ref{NutationRotaryEcho}(a), we see that the rotary echo data closely approximates the simulated Hahn echo data ($S^\text{P}(t)$, as described in Appendix \ref{Numerics}) when the Hahn echo data are scaled by 2 in time. Note that the rotary echo data also last much longer than the nutation data, indicating that the decay from $H_1$ inhomogeneity is not negligible. In order to isolate the decay caused specifically by the $H_1$ field inhomogeneity across the sample, we divide the nutation data by the simulated, scaled Hahn echo data, effectively removing the component of the decay caused by $-(1/2)\mathcal{H}^\text{P,P}_{xx}$. We use the Hahn echo data rather than the rotary echo data since it lasts slightly longer than the rotary echo data; thus, we ascribe more of the overall nutation curve decay to the $H_1$ inhomogeneity (representing a ``worst-case'' scenario for $H_1$ inhomogeneity across the sample). While this procedure produces noise near the tail of the decay, the fits discussed below are largely unaffected because the noise is random [Fig. \ref{NutationRotaryEcho}(a)].

 Next, we try to infer an $H_1$ probability distribution $p(\gamma H_1 / 2\pi)$ which could cause the remaining decay in the altered nutation experiment. Based upon previous work \cite{Li2008}, we assume that a sum of two Gaussians is a reasonable approximation to the shape of the $H_1$ distribution. This allows us to write an analytical time-domain decay function, which we fit to the nutation data (as altered above), with good results [Fig. \ref{NutationRotaryEcho}(a), insets]. The parameters from the fit determine the shape of $p(\gamma H_1 / 2\pi)$ [Fig. \ref{NutationRotaryEcho}(b)], which itself provides a measure of the pulse imperfection (a spread in actually applied pulse angles).

 Finally, we incorporate the spread in $\epsilon$ into a modification of the original ``product of cosines'' decay model: $ \cos^N(\epsilon) \rightarrow \sum_i p_i \cos^N(\epsilon_i)$ for a range of angles $\epsilon_i=|\theta_i - \pi|$ with probabilities $p_i$. The corrected decay model still decays much more slowly than the DTC oscillation at $\theta = \pi$, even when we include the effect of the phase transients as described [Fig. \ref{NutationRotaryEcho}(c)]. Thus, this ``worst-case-scenario'' effect from the $H_1$ inhomogeneity across the sample is insufficient to account for the observed decay of the DTC oscillations at $\theta=\pi$.

\subsection{Studying the effect of the internal Hamiltonian during a nonzero-duration pulse}

\begin{figure}
\includegraphics[width=0.5\textwidth]{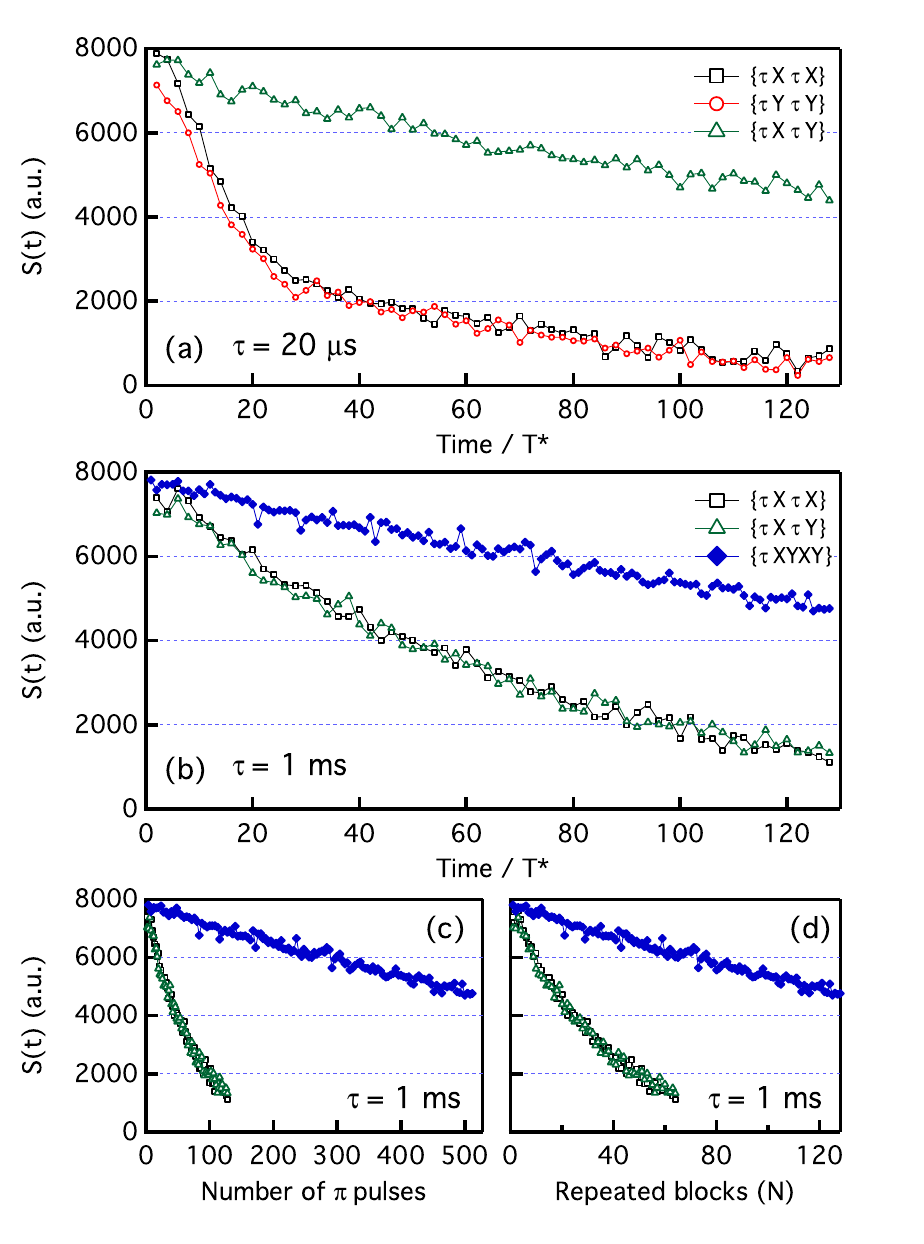}%
\caption{\label{NonzeroDurationPulses}(a) Significant differences in the decay rate between sequences that are identical in the delta-function $\pi$-pulse approximation. At $\tau=20$\,$\mu$s, the pulse sequences $\{X,X\}$ (black open squares) and $\{Y,Y\}$ (red open circles) produce very different lifetimes than $\{X,Y\}$ (green open triangles). The effect of the internal Hamiltonian during the pulse time $t_p$ creates differences between these sequences, which gives the latter sequence a much longer lifetime (see text). Because the signal is only observed every two cycles, the oscillations in the signal are not seen here. (b) Results of the pulse sequence $\{\tau - X_\pi - Y_\pi - X_\pi - Y_\pi\}^N$ (closed blue diamonds), which again exhibits an extended lifetime compared to the original DTC sequence, even at long $\tau$. (c), (d) The difference in lifetimes as a function of absolute time is significant, but displaying the pulse sequences as functions of the number of applied $\pi$ pulses or repeated blocks shows even more dramatic differences. Here, we define $T^*$ as the shortest repeated period, ignoring the phase of the pulses. For $\{\tau - X_\pi - Y_\pi - X_\pi - Y_\pi\}^N$, $T^*=\tau + 4t_p$, while for $\{X,Y\}$, $T^*=\tau + t_p$.}
 \end{figure}

  Since these experimental causes have been shown to be too small to account for the observed decay at $\theta = \pi$, we return to the approximation that the applied pulses are zero duration (delta-function) pulses. To study the effect of the internal Hamiltonian during a $\theta = \pi$ pulse of nonzero duration, we implement modified versions of the DTC sequence with different sets of pulse phases, since this allows us to selectively manipulate the effective internal Hamiltonian during the pulse. Defining $\{\alpha,\beta\}\equiv\{\tau - \alpha_\pi - \tau - \beta_\pi\}^N$, we compare the sequences $\{X,X\}$, $\{Y,Y\}$, and $\{X,Y\}$, which should produce identical results for zero-duration $\pi$ pulses. However, Fig. \ref{NonzeroDurationPulses}(a) shows that the signal from $\{X,Y\}$ lasts far longer than the signal from either $\{X,X\}$ or $\{Y,Y\}$ for short $\tau$, demonstrating that the non-zero pulse duration plays an important role in the observed decay.

  These results may be qualitatively explained for pulses of non-zero duration, when we use the identity $\mathcal{H}^\text{P,P}_{xx} + \mathcal{H}^\text{P,P}_{yy} + \mathcal{H}^\text{P,P}_{zz} = 0$ (used, e.g., in the WAHUHA sequence to average the total dipolar evolution to zero \cite{Waugh1968, Haeberlen1976}). The sequence $\{X,Y\}$ has average Hamiltonian $\mathcal{H}^{(0)}=2\mathcal{H}^\text{P,P}_{zz}\tau - (\mathcal{H}^\text{P,P}_{xx} + \mathcal{H}^\text{P,P}_{yy})t_p/2 = (2\tau + t_p/2)\mathcal{H}^\text{P,P}_{zz}$, and will thus leave the original state unaffected to zeroth order in the Magnus expansion. This is in contrast to $\{X,X\}$, which has average Hamiltonian $\mathcal{H}^{(0)}=2\mathcal{H}^\text{P,P}_{zz}\tau - \mathcal{H}^\text{P,P}_{xx} t_p$.

  The approximate average Hamiltonian analysis explaining this result breaks down when $\tau$ is long \cite{Haeberlen1968, Haeberlen1976}, where the advantage of $\{X,Y\}$ over $\{X,X\}$ is lost [Fig. \ref{NonzeroDurationPulses}(b)]. Although we cannot rely on the convergence of the Magnus expansion at long $\tau$, we can still try to extend the decay envelope by moving even farther from the original DTC sequence and applying pulse sequences which use a burst of $\pi$ pulses instead of one \cite{Dong2008}. In Fig. \ref{NonzeroDurationPulses}(b), we show the results of $\{\tau - X_\pi - Y_\pi - X_\pi - Y_\pi\}^N$, which again shows an extended lifetime, even at long $\tau$. The signal resulting from this sequence lasts longer than the original DTC sequence even in absolute time, despite the increased number of necessarily imperfect pulses [Fig. \ref{NonzeroDurationPulses}(b)-\ref{NonzeroDurationPulses}(d)].

  The analysis in this section pertains to $\theta = \pi$. When $\theta$ is adjusted away from $\pi$, the effect of the interactions during the pulse should grow, as terms that were strictly zero at $\theta = \pi$ begin to turn on \cite{Li2007,Li2008}. Thus, we expect the dipolar interactions during the pulse to produce a decay envelope at $\theta \neq \pi$ which will limit the echoes shown in Sec. \ref{Refocusing}. Creating echoes that are able to rise above this envelope will be difficult, since it is harder to undo the many different terms which arise for $\theta \neq \pi$, but it may be possible.

 \section{CONCLUSION}

 We became especially interested in descriptions of DTC phenomena when reports appeared in the literature of period doubling in driven systems, since comparable behavior emerged for long cycle times in our prior studies of periodically driven NMR systems \cite{Li2007, Li2008}. From these studies, we developed a model that took into account the interactions during $\pi$ pulses \cite{Li2007,Li2008}, which we put to good use in the small cycle time limit \cite{Dong2008, Frey2012}. However, most of our originally puzzling data \cite{Li2008} lie beyond the reach of our model, since it relied on the Magnus expansion, which diverges for long cycle times \cite{Haeberlen1968, Haeberlen1976}. Thus, we wondered if the growing theoretical framework around DTC order could shed light on our still unexplained results, and we began to conduct similar experiments to the ones which had been published for systems of trapped ions \cite{Zhang2017} and diamond NV centers \cite{Choi2017}. 

 Both this system and the system of diamond NV centers are very different from the system of trapped ions, being large systems with long-ranged dipolar couplings in three dimensions. The ADP crystal studied here is itself strikingly different from the system of NV centers, being a dense, organized crystal with no significant sources of disorder. Nevertheless, despite the many differences in the spin Hamiltonian for our system, our results are strikingly similar to the results achieved in both of these prior DTC experiments. Furthermore, our experiment allows us to explore a very large region in the $(\theta,\tau)$ parameter space, where we observe robust DTC oscillations across a remarkably broad range in $\tau$; in particular, $0.03 < W^\text{P,P}\tau < 3200$\,radians.

 The clean spatial crystal studied here should be even less conducive to MBL than the systems in prior experiments \cite{Ho2017,Nandkishore2015}; if MBL plays a role in our experiments, that would seem to require MBL to occur in highly unanticipated regimes. A prethermal DTC state could explain the observations of persistent DTC oscillations like the ones observed here. However, for our system $\braket{H_{zz}^\text{P,P}(t=0)}=0$, which suggests that the initial state is at an infinite temperature relative to the effective Floquet Hamiltonian. This seems to rule out a prethermal explanation for our observations, since that normally requires the system to start below some finite critical temperature \cite{Else2017}.

 The decay envelope of the observable DTC oscillations in our system was bounded by a simple ``product of cosines'' dephasing model for certain values of $\theta$ and $\tau$; however, using the DTC echo as a new probe of the state shows us that the density matrix produced by the DTC sequence retains a coherent memory of its initial state.

 Turning to the decay envelope of the DTC oscillations at $\theta=\pi$, we see clear evidence of the effect of $\mathcal{H}_\text{int}$ during nonzero duration pulses. We suggest that more in-depth studies of the DTC lifetime should account for the action of terms in the internal Hamiltonian during a pulse, since these small terms can have significant effects over the course of many repeated pulses.

Driven, out of equilibrium many-body systems are thought to be interesting hunting grounds for new physics and phases of matter. Solid-state NMR can aid in this search, by exploiting the large separation between $T_1$ and $T_2$, the ability to edit the effective Hamiltonian using pulses, and other tricks in the NMR toolbox.

\textit{Note added}. Recently, the authors of an interesting related experiment contacted us, alerting us to their liquid state NMR search for temporal order of periodically-driven spins in star-shaped clusters \cite{Pal2018}.  They study a unique spin Hamiltonian, and they explore a range of cluster sizes (with $N=1, 4, 10$, and 37 spins).
 
  
  \section*{ACKNOWLEDGMENTS}
      We thank C. W. von Keyserlingk, V. Khemani, C. Nayak, N. Yao, and M. Cheng for helpful discussions. We also thank C. Grant and D. Johnson for help in constructing the NMR probe, K. Zilm for recommending the ADP sample, and S. Elrington for assistance with implementing cross polarization. This material is based upon work supported by the National Science Foundation under Grant No. DMR-1610313. R.L.B. acknowledges support from the National Science Foundation Graduate Research Fellowship under Grant No. DGE-1122492. 

 \appendix

\section{\label{CrystalSymmetry}SYMMETRIES OF CERTAIN SUBLATTICES IN ADP CRYSTAL}

Ammonium dihydrogen phosphate is a tetragonal crystal with unit cell dimensions $a=b=7.4997$\,\angstrom, $c=7.5494$\,\angstrom. Here we show that the secular dipolar coupling for the $^{31}$P and $^{14}$N sublattices of the $I\overline{4}2d$ ADP crystal are invariant under shifts to any other site of the sublattice. We do so explicitly by writing the coordinates of the $^{31}$P sublattice relative to the unit cell \cite{Khan1973, West1930}:
\begin{equation}
  (a,b,c) = \{(0, 0, 0), (\tfrac{1}{2}, \tfrac{1}{2}, \tfrac{1}{2}), (\tfrac{1}{2}, 0, \tfrac{1}{4}), (0, \tfrac{1}{2}, \tfrac{3}{4})\},
\end{equation}

with $^{14}$N sites $(a,b,c+\tfrac{1}{2})$. Simply, we translate each coordinate to the origin, and examine the symmetry. These positions go into themselves by translations $(\tfrac{1}{2}, \tfrac{1}{2}, \tfrac{1}{2})$, but translations by $(-\tfrac{1}{2}, 0, -\tfrac{1}{4})$ or $(0, -\tfrac{1}{2}, -\tfrac{3}{4})$ produce a set of coordinates $(a', b', c') = (a, b, -c)$, inverted in $c$. However, the crystal is symmetric under rotations about $c$ by $180^{\circ}$ ($=360^{\circ}/2$, 2 symmetry) such that $(a',b',c') \rightarrow -(a,b,c)$, e.g., a complete inversion. Thus, for any coordinate vectors $\vec{r}_A$ and $\vec{r}_B$ of any $^{31}$P or $^{14}$N nucleus, the internuclear vector transforms as $\vec{r}=(\vec{r_A} - \vec{r_B}) \rightarrow -\vec{r}$. Relative to the $z$ axis as defined by the external $B$ field (not necessarily along $c$), we then have $\cos (\theta) = \vec{r}\cdot \vec{B} / (|\vec{r}||\vec{B}|) \rightarrow -\cos (\theta)$, so that $\cos ^2(\theta)$ is invariant. Since all distances are preserved, the invariance of $B_{ij}(r,\theta)$ follows. These arguments may be immediately extended to the ammonium $^1$H, whose average positions reside on the nitrogen sites.

For the acid $^1$H, these symmetry arguments only hold for particular orientations of the crystal relative to the static field. To see this, note that the average positions of these eight $^1$H are
\begin{align}
  &(0,0,0),(\tfrac{1}{2},\tfrac{1}{2},\tfrac{1}{2}) \nonumber \\
  &+ \{(x, \tfrac{1}{4}, \tfrac{1}{8}), (-x, \tfrac{3}{4}, \tfrac{1}{8}), (\tfrac{1}{4}, -x, \tfrac{7}{8}), (\tfrac{3}{4}, x, \tfrac{7}{8})\},
\end{align}
with $x=0.147$. Upon translation by $(-\tfrac{1}{2}, 0, -\tfrac{1}{4})$ or $(0, -\tfrac{1}{2}, -\tfrac{3}{4})$, rather than being invariant after inversions in $c$, the lower-symmetry locations of the acid $^1$H are invariant only after a $180^{\circ}$ rotation about either $a$ or $b$. This results in the transformed unit-cell coordinates $(a',b',c') = (a,-b, -c)$ or $(-a,b,-c)$, neither of which preserves $\vec{r}\cdot \vec{B}$ for internuclear vectors $\vec{r}$ in general. Nonetheless, the $\{(0, 0, 0), (\tfrac{1}{2}, \tfrac{1}{2}, \tfrac{1}{2})\}$ and $\{(\tfrac{1}{2}, 0, \tfrac{1}{4}), (0, \tfrac{1}{2}, \tfrac{3}{4})\}$ $^{31}$P sublattices independently maintain identical sets of coupling constants to the acid $^1$H, which become the same under certain orientations of the crystal relative to the external field. Specifically, if the strong external field lies purely in the $x$-$z$ or $y$-$z$ planes (e.g. either $B_x=0$ or $B_y=0$ relative to the crystal axes), the $^{31}$P couplings to the acid $^1$H will be identical for each $^{31}$P nucleus. Since the azimuthal angle of $H_0$ with respect to the crystal axes $(a,b,c)$ is $\phi \approx m90^{\circ}, m=0,1,2,3$ in our experiment (see main text), we see this symmetry in the numerics for the orientation angle which best approximates the data (Fig \ref{LineShape}) \footnote{We note that these symmetries will also extend to other groups of spin-spin couplings, for instance the $^{14}$N-$^{14}$N coupling.}.

\section{\label{Numerics}NUMERICS OF THE $^{31}$P SPIN HAMILTONIAN IN ADP CRYSTAL}

We simulate a lattice of spins with published atomic positions, with two modifications appropriate for the motionally-narrowed NMR spectrum: (1) we locate the acid $^1$H in time-averaged positions halfway between the nearest PO$_4$ oxygens, and (2) we locate ammonium $^1$H in time-averaged positions at the nitrogen lattice sites \cite{Khan1973, West1930}. These modifications account for motions that are very rapid compared to NMR timescales. We begin by treating each $^{31}$P location in the unit cell in turn as the origin of a large cluster of spins on the lattice, only including spins within a radius of $R\approx20.25\angstrom$ around the origin (corresponding to 325 $^{31}$P, 322 $^{14}$N, and 1932 $^1$H). We then calculate the line shapes from $\mathcal{H}_{zz}^\text{P,P}$, $\mathcal{H}_{zz}^\text{P,H}$, and $\mathcal{H}_{zz}^\text{P,N}$ separately; in order to calculate the line shape resulting from, e.g., $\mathcal{H}_{zz}^\text{P,H}$, we only calculate the coupling constants $B^\text{H}_{1j}$ between the central $^{31}$P spin and all $^1$H spins in the cluster, and use these $B^\text{H}_{1j}$ values for our simulation. We do this for a given sample orientation, parameterized by the respective azimuthal and polar angles $(\theta_c, \phi_c)$ of the static field $H_0$ relative to the crystal axes, $(x,y,z)=(a,b,c)$. For each pair of spins, we first approximate the coupling as $\sum B_{1j} (2I_{z_1}I_{z_j})$, since this is an analytically solvable model \cite{Lowe1957, Sorensen1984}. For an initial density matrix proportional to $I_{y_T}$, the signal measured for a single spin-$\tfrac{1}{2}$ coupled to spin-``$s$'' evolves as $S(t)=\braket{I_{y_1}(t)}/\braket{I_{y_1}(0)}=\prod_j\sum_k p_k \cos[m_k (2B_{1j})t/\hbar]$, where $m_k$ are the possible $m_z$ quantum numbers for a spin-$s$ particle, and $p_k$ are the corresponding probabilities (e.g. $m_k=\{+1,0,-1\}$ and $p_k=\{\tfrac{1}{3},\tfrac{1}{3},\tfrac{1}{3}\}$ for spin-1).

\begin{figure}
\includegraphics[width=0.45\textwidth]{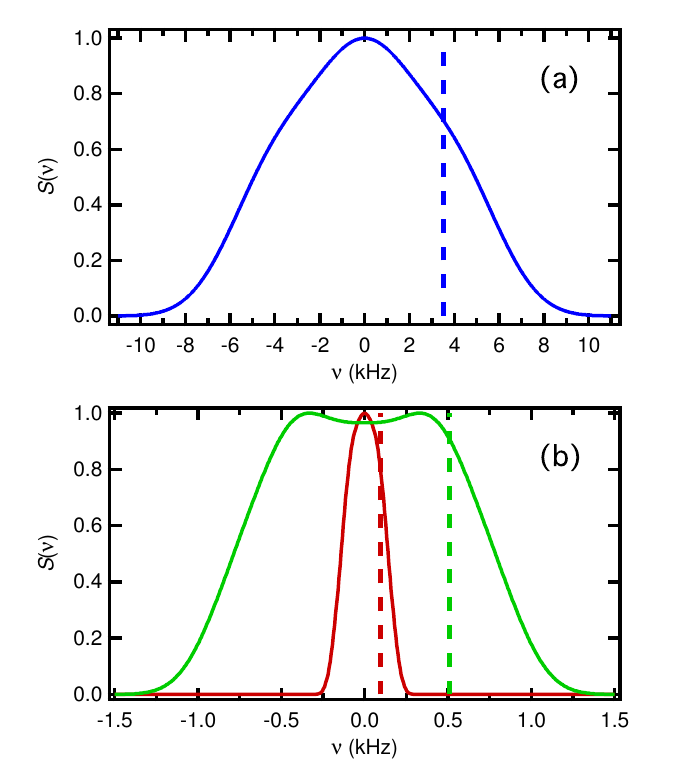}%
\caption{\label{ComputationalSpectra} Computed $^{31}$P spectra $S(\nu)$, with marked rms frequencies $W/2\pi$. (a) The pure $^{31}$P-$^1$H dipolar spectrum $S^\text{P,H}(\nu)$ (solid line), with marked rms frequency (dashed line) $W^\text{P,H}/2\pi=3500$\,Hz. (b) The pure $^{31}$P-$^{14}$N dipolar spectrum $S^\text{P,N}(\nu)$ (red solid line) is narrower than the pure $^{31}$P-$^{31}$P dipolar spectrum $S^\text{P,P}(\nu)$ (green solid line). The rms frequencies (dashed lines) $W^\text{P,N}/2\pi=97$\,Hz and $W^\text{P,P}/2\pi=508$\,Hz are marked. Not shown here are $S^\text{P,PN}(\nu)$ [similar to Fig. \ref{LineShape}(c), red circles] with rms frequency $W^\text{P,PN}/2\pi=517$\,Hz, and $S^\text{P,HPN}(\nu)$ [see Fig. \ref{LineShape}(b), blue squares] with rms frequency $W^\text{P,HPN}/2\pi=3538$\,Hz.}
\end{figure}

 We define $S^\text{P},S^\text{H}$, and $S^\text{N}$ to be the signals calculated from a spin Hamiltonian containing only $\mathcal{H}_{zz}^\text{P,P}, \mathcal{H}_{zz}^\text{P,H}$, or $\mathcal{H}_{zz}^\text{P,N}$, respectively, then calculate the magnetization decay using the appropriate spin values in the formula above. For the $^{31}$P-$^{31}$P coupling, we have (recalling that $^{31}$P has a spin-$\tfrac{1}{2}$ nucleus)
\begin{align}
 S^\text{P}(t) &= \prod_j \cos(\frac{3}{2}B^\text{P}_{1j}t/\hbar),
\end{align}
where we include a factor of $3/2$ in the coupling constant to account for the difference between the $3I_{z_i}I_{z_j}$ in the full dipolar coupling for ``like spins'' and the $2I_{z_i}I_{z_j}$ in our Ising-type model for ``unlike spins.'' For couplings of $^{31}$P to the spin-$\tfrac{1}{2}$ $^{1}$H, we have
\begin{align}
S^\text{H}(t) &= \prod_j \cos(B^\text{H}_{1j}t/\hbar).
\end{align} 
Finally, for couplings to the spin-1 $^{14}$N nuclei, we have
\begin{align}
S^\text{N}(t) &= \prod_j \frac{1}{3}\{2\cos[(2B^\text{N}_{1j})t/\hbar]+1\}.
\end{align}
To arrive at these $S(t)$, we have chosen one of the four unique lattice positions of $^{31}$P in the unit cell to serve as the origin. We repeat this procedure with the lattice centered at each of the four unique $^{31}$P positions in the crystal structure, and average the four results to arrive at a total time-domain signal. To calculate the combined effect of multiple interactions (e.g., include $^{31}$P-$^{31}$P, $^{31}$P-$^1$H, and $^{31}$P-$^{14}$N interactions), we multiply the corresponding time data [e.g., $S^\text{P,HPN}(t)=S^\text{P,H}(t) \times S^\text{P,P}(t) \times S^\text{P,N}(t)$]. We do a complex Fourier transform of $S(t)$ to produce a spectrum $S(\nu)$ (e.g., Fig. \ref{ComputationalSpectra}), from which we derive a mean square coupling strength $(W/2\pi)^2 = \braket{\nu^2} = \sum_\nu\nu^2\text{Re}[S(\nu)]/\sum_\nu \text{Re}[S(\nu)]$.

Lastly, we can compare these $W$ to the rms $B_{1j}$ values themselves, after proper weighting. For $^{31}$P-$^{31}$P, we find $W^\text{P,P} = \frac{3}{2}B_{\text{rms}}^\text{P}/\hbar$, where we include $3/2$ for the reasons discussed above. For $^{31}$P-$^{1}$H, we find $W^\text{P,H} = B_{\text{rms}}^\text{H}/\hbar$. For $^{31}$P-$^{14}$N, we find $W^\text{P,N} = 2\sqrt{2/3}B_{\text{rms}}^\text{N}/\hbar$, where we have again used the spin statistics for the spin-1 $^{14}$N:   $(\hbar W^\text{P,N})^2 = \sum_k p_k(m_k2B^\text{N}_\text{rms})^2$, with $p_k=\{\tfrac{1}{3},\tfrac{1}{3},\tfrac{1}{3}\}$ and $m_k=\{+1,0,-1\}$.

\section{\label{WindowDependence}DEPENDENCE OF THE CRYSTALLINE FRACTION ON THE WINDOW SIZE USED BY THE FOURIER TRANSFORM}

\begin{figure}
\includegraphics[width=0.45\textwidth]{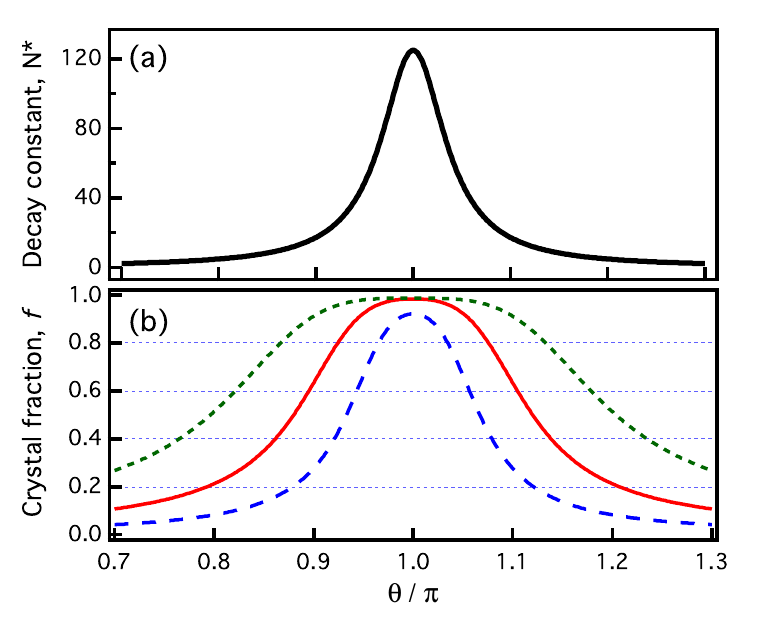}%
\caption{\label{WindowEffect}(a) Lorentzian decay constant $N^*$ as a function of $\theta$ used in this example. (b) Crystalline fraction calculated for the distribution in (a), using $N=1$--$128$ (blue dashed line), $N=1$--$50$ (red solid line), and $N=1$--$20$ (green dotted line).}
\end{figure}

When we used fewer points in our FT window, e.g., $N=51$--$100$ of $S(t)$, the crystalline fraction $f(\theta)$ acquired flatter regions around $\theta = \pi$, fitting better to super-Gaussians than Gaussians. A simple model shows that this arises from the definition of the crystalline fraction. We model a signal which oscillates under an exponential decay $S(N) = (-1)^N \exp(-N/N^*)$, where the decay constant $N^*$ depends on $\theta$. In this model, we use
\begin{equation}
  N^*(\theta) = 125 \frac{0.04^2}{(\theta/\pi-1)^2 + 0.04^2}
\end{equation}
as shown in [Fig. \ref{WindowEffect}(a)]; the Lorentzian dependence of $N^*$ on $\theta$ is a reasonable description of much of our data. Using this $N^*(\theta)$, Fig. \ref{WindowEffect}(b) shows the calculated crystal fraction $f$ using three different Fourier transform window sizes: $N=1$--$128$, $N=1$--$50$, and $N=1$--$20$. The change in the window size is sufficient to produce flatter tops; the crystal fraction data shown in Figs. \ref{GaussiansCW}-\ref{SuperGaussians} should be read with this in mind. 

%

\end{document}